\pdfoutput=1

\documentclass[11pt,twoside,a4paper,cmspaper,final,collab]{cms-tdr}
\def\svnVersion{494657P}\def\cmsCernNoTag{CERN-EP-2018-207}\def\cmsCernDate{\today}\def\cmsMessage{Published in Physics Letters B as \href{http://dx.doi.org/10.1016/j.physletb.2019.03.064}{\doi{10.1016/j.physletb.2019.03.064.}}}

\begin{document}\cmsNoteHeader{HIG-17-013}

\hyphenation{had-ron-i-za-tion}
\hyphenation{cal-or-i-me-ter}
\hyphenation{de-vices}
\RCS$HeadURL: svn+ssh://svn.cern.ch/reps/tdr2/papers/HIG-17-013/trunk/HIG-17-013.tex $
\RCS$Id: HIG-17-013.tex 488111 2019-02-02 17:32:52Z smgascon $
\newlength\cmsFigWidth
\newlength\cmsTabSkip\setlength{\cmsTabSkip}{1ex}
\ifthenelse{\boolean{cms@external}}{\setlength\cmsFigWidth{0.98\columnwidth}}{\setlength\cmsFigWidth{0.6\textwidth}}
\ifthenelse{\boolean{cms@external}}{\providecommand{\cmsLeft}{upper\xspace}}{\providecommand{\cmsLeft}{left\xspace}}
\ifthenelse{\boolean{cms@external}}{\providecommand{\cmsRight}{lower\xspace}}{\providecommand{\cmsRight}{right\xspace}}

\ifthenelse{\boolean{cms@external}}{\providecommand{\cmsTable}[1]{#1}}{\providecommand{\cmsTable}[1]{\resizebox{\textwidth}{!}{#1}}}
\newcommand{\mH}{\ensuremath{m_{\mathrm{H}}}\xspace}
\newcommand{\mgg}{\ensuremath{m_{\gamma\gamma}}\xspace}

\cmsNoteHeader{HIG-17-013}
\title{Search for a standard model-like Higgs boson in the mass range between 70 and 110\GeV in the diphoton final state in proton-proton collisions at $\sqrt{s}=8$ and 13\TeV}

\date{\today}

\abstract{The results of a search for a standard model-like Higgs boson in the mass range between 70 and 110\GeV decaying into two photons are presented.  The analysis uses the data set collected with the CMS experiment in proton-proton collisions during the 2012 and 2016 LHC running periods.  The data sample corresponds to an integrated luminosity of 19.7 $(35.9)$\fbinv at $\sqrt{s}=8$ (13)\TeV.  The  expected and observed 95\% confidence level upper limits on the product of the cross section and branching fraction into two photons are  presented. The observed upper limit for the 2012\,(2016) data set ranges from 129\,(161)\unit{fb} to 31\,(26)\unit{fb}.  The statistical combination of the results from the analyses of the two data sets in the common mass range between 80 and 110\GeV yields an upper limit on the product of the cross section and branching fraction, normalized to that for a standard model-like Higgs boson, ranging from 0.7 to 0.2, with two notable exceptions: one in the region around the \PZ boson peak, where the limit rises to 1.1, which may be due to the presence of Drell--Yan dielectron production where electrons could be misidentified as isolated photons, and a second due to an observed excess with respect to the standard model prediction, which is maximal for a mass hypothesis of 95.3\GeV with a local (global) significance of 2.8\,(1.3) standard deviations.}

\hypersetup{%
pdfauthor={CMS Collaboration},%
pdftitle={Search for a standard model-like Higgs boson in the mass range between 70 and 110 GeV in the diphoton final state in proton-proton collisions at sqrt(s)= 8 and 13 TeV},%
pdfsubject={CMS},%
pdfkeywords={CMS, physics, Higgs, diphoton}}

\maketitle

\section{Introduction}
\label{sec:intro}
Within the standard model (SM) of particle physics~\cite{SMHiggsO,SMHiggsT,SMHiggsTh},
particle masses arise from the spontaneous breaking of  electroweak symmetry,
which is
achieved through the Brout--Englert--Higgs mechanism~\cite{HiggsExpO,HiggsExpT,HiggsExpTh,PhysRevLett.13.585,
PhysRev.145.1156,PhysRev.155.1554}. In its minimal version,
electroweak symmetry breaking is realized through the introduction of a doublet of complex scalar fields.
At the end of the process, only one scalar field
remains and the corresponding quantum, the Higgs boson, should be experimentally observable.  In 2012, both the ATLAS~\cite{Aad:2012tfa} and
CMS~\cite{Chatrchyan:2012xdj,Chatrchyan:2013lba} Collaborations observed a new boson with
a mass of approximately 125\GeV whose properties are at present compatible with those of the
SM Higgs boson.
The analyses of data in the diphoton final state leading to this discovery probed an invariant mass range extending
from 110
to 150\GeV.

However,
physics beyond the SM (BSM) can also provide a Higgs boson that is compatible with the observed 125\GeV boson.
The extended parameter space of
several BSM models,
for example generalized
models containing two Higgs doublets (2HDM)~\cite{Celis:2013rcs,Coleppa:2013dya,Chang:2013ona,Bernon:2015wef,Cacciapaglia2016} and the next-to-minimal supersymmetric model (NMSSM)~\cite{Fayet:1974pd,Barbieri:1982eh,Dine:1981rt,Nilles:1982dy,Frere:1983ag,Derendinger:1983bz,PhysRevD.39.844,Drees:1988fc,Ellwanger:1993xa,Ellwanger:1995ru,Ellwanger:1996gw,Elliott:1994ht,King:1995vk,Franke:1995tc,Maniatis:2009re,Ellwanger:2009dp,1674-1137-38-7-073101,Ellwanger:2015uaz,Guchait:2016pes,Cao:2016uwt}, gives rise to a rich and interesting phenomenology, including the presence of
additional Higgs bosons, some of which could have masses below 125\GeV.
Such models provide good motivation for
extending searches for
Higgs bosons
to masses
as far
below
$\mH=110\GeV$ as possible, where \PH refers to an additional Higgs boson which is ``SM-like", meaning that the relative contributions of the production processes are
similar to those of the SM.

  The $\HGG$ decay channel provides a clean final-state topology
that allows the mass of
a Higgs boson in the search range to be reconstructed with high precision.
The primary production mechanism
for Higgs bosons in proton-proton (\Pp\Pp) collisions
at
the CERN LHC
is gluon fusion ($\Pg\Pg\PH$), with additional smaller contributions from vector boson fusion (VBF) and production in association with a $\PW$ or $\PZ$ boson ($\mathrm{V}\PH$), or with a \ttbar pair ($\ttbar\PH$). The dominant sources of background are
irreducible direct diphoton production, and the reducible $\PPTOGAMJET$ and \Pp\Pp $\to$ jet $+$ jet processes, where the jets are misidentified as isolated photons. An additional source of reducible background relevant for the search range below
$\mH=110\GeV$ is
Drell--Yan dielectron production, where electrons could be misidentified as isolated photons.

The CERN LEP collaborations~\cite{Barate:2003sz}, in the context of the search for the SM Higgs boson, explored the mass range below 110\GeV extensively
in the $\mathrm{V}\PH$ production modes,
in the \bbbar and \TT
channels. Several of the BSM models mentioned above predict reduced decay rates in these channels with respect to SM predictions
and enhanced decay rates in the diphoton channel.  The "low-mass" search in the diphoton decay channel by 
ATLAS~\cite{PhysRevLett.113.171801}, performed in the mass range of $65<\mgg<110\GeV$ at a center-of-mass energy of 8\TeV, 
found no significant excess with respect to
expectations.

This
letter presents
the result of a search in the diphoton channel for an
additional
Higgs boson
with an invariant mass lower than 110\GeV,
whose natural width is small compared to the detector resolution.
The search is performed on
a data set collected in 2012 and 2016
with the
CMS detector at the
LHC, corresponding to, respectively,
integrated luminosities of 19.7\fbinv at a center-of-mass energy of 8\TeV,
referred
to as the ``8\TeV data", and 35.9\fbinv at 13\TeV, the
"13\TeV data".

The analysis is based on a search for a localized excess
in the diphoton invariant mass spectrum over a smoothly falling background
from prompt diphoton production and from events with at least one jet misidentified as a photon, in addition to the Drell--Yan contribution. It uses an
extended version of the method developed
by the CMS Collaboration for the observation and the measurement of the properties of the 125\GeV
boson~\cite{HIG-13-001,
CMS-PAS-HIG-16-040}.  The invariant mass range explored in the 8
(13)\TeV data is 80 $(70)<\mgg<110\GeV$. The principal challenges associated with a search in the diphoton decay channel in this mass range are the ability to trigger on events while maintaining acceptable rates, and the background from
\PZ bosons decaying to electron pairs that, through misidentification, could appear to result in two isolated photons.  To achieve the best possible sensitivity, the events are separated into classes.
Multivariate analysis (MVA) techniques are used both for photon identification and event classification, and
the signal is extracted from the background using a fit to the diphoton mass spectrum in all event classes.

\section{The CMS detector}
\label{sec:det}

A detailed description of the CMS detector, together with a definition of the coordinate system used and the relevant kinematic variables, can be found in Ref.~\cite{Chatrchyan:2008zzk}.  The central feature of the CMS
apparatus is a superconducting solenoid of 6\unit{m} internal diameter,
 providing
a magnetic field of 3.8\unit{T}.
 Within the
solenoid volume are a silicon pixel and strip tracker, a lead tungstate crystal electromagnetic calorimeter (ECAL),
 and a brass and scintillator hadron calorimeter (HCAL),
 each composed of a barrel and two endcap sections.
 Forward calorimeters extend the pseudorapidity ($\eta$)
coverage provided by the barrel and endcap detectors.
Muons are
detected in gas-ionization
chambers embedded in the steel flux-return yoke outside the solenoid.
The
ECAL, surrounding the tracker volume, consists of
75\,848 lead tungstate crystals, which provide coverage in
$\abs{ \eta }< 1.48 $ in a barrel region (EB) and $1.48 <\abs{ \eta } < 3.0$ in two endcap regions (EE).
Preshower detectors consisting of two planes of silicon sensors interleaved with a total of $3 X_0$ of lead are located in front of each EE detector.
In
the EB, an energy resolution of about 1\% is achieved for unconverted or late-converting photons that have energies in the range of
tens of \GeV.
For the remaining barrel photons,
a resolution of about 1.3\% is achieved up to
$\abs{\eta}=1$, rising to about 2.5\% at $\abs{\eta}=1.4$. In the
EE, an energy
resolution
for unconverted or late-converting photons
of about 2.5\% is achieved, while for the remaining endcap photons it is
between 3 and 4\%~\cite{CMS:EGM-14-001}.
\section{Measurement of the diphoton mass spectrum}
\subsection{Trigger and simulation}
\label{sec:triggerSim}
Events of interest are selected using a two-tiered trigger system~\cite{Khachatryan:2016bia}. The first level,
composed of custom hardware processors, uses information from the calorimeters and muon detectors to select events at a rate of around 100\unit{kHz} within a time interval of less than 4\unit{\mus}. The second level, known as the high-level trigger (HLT), consists of a farm of processors running a version of the full event reconstruction software optimized for fast processing, and reduces the event rate to less than 1\unit{kHz} before data storage. For this analysis,
diphoton HLT paths with asymmetric transverse momentum
($\pt$) thresholds are used. 

In the case of the 8\TeV data, the same paths are used as in~\cite{HIG-13-001}.
The paths that select
almost all of the events
impose thresholds of 26 and 18\GeV on the $\pt$ of the individual photon trigger objects, and
minimum requirements on the invariant mass of diphoton trigger objects of either 60 or 70\GeV depending on the data-taking period.

For the 13\TeV data, two dedicated HLT paths are used, both with photon $\pt$ thresholds of 30 and 18\GeV.  One path has nearly identical requirements to those used in~\cite{CMS-PAS-HIG-16-040},
except that only events with both photon candidates
in the EB are selected.
This path requires each of the photon candidates to satisfy criteria on the ratio of its energy in the HCAL and in the ECAL (H/E), and on either
shower shape or on its isolation energy.
The other path selects events with photon candidates from any part of the ECAL,
but they must satisfy more stringent shower shape requirements
as well as the requirements on
both isolation energy and
H/E.
In addition, both paths impose
a veto on the presence of hits compatible with the photon direction in the silicon pixel detector, and
require that the invariant mass of the two photon candidates be greater than 55\GeV. 

These requirements
limit the search range
to $\mgg>70$ $(80)\GeV$ for the 13\,(8)\TeV data, in order to avoid the portion of the offline diphoton spectrum that is distorted due to turn-on effects from the HLT criteria. For both data sets, the trigger efficiency is measured from $\PZ\to\EE$ events using
the tag-and-probe technique~\cite{tagNprobe}, except for the pixel hit veto requirement relevant for the triggering of the 13\TeV data,
where the efficiency is measured using diphoton events in data
that have passed the trigger used in~\cite{CMS-PAS-HIG-16-040}, which does not require a pixel veto.

Monte Carlo
(MC) simulations are used to produce
SM Higgs boson events from
all production processes
($\Pg\Pg\PH$, VBF, $\mathrm{V}\PH$, and $\ttbar\PH$), with
invariant masses ranging from 70 to 110\GeV.
These events are the input to the signal modeling procedure,
representing a new resonance decaying to two photons. In the case of the 8\TeV data, for the
$\Pg\Pg\PH$ and VBF processes, these events are generated at next-to-leading order (NLO) in perturbative quantum chromodynamics (QCD) using \POWHEG 1.0~\cite{powheg3,powheg4,powheg5,powheg1,powheg2}, while the
events from the associated production processes are generated at leading order (LO) with \PYTHIA 6.426~\cite{PYTHIA}.  For the 13\TeV data,
events are generated at NLO using \MGvATNLO 2.2.2~\cite{Alwall:2014hca} with FxFx merging~\cite{FXFX}, for all  production processes.
Events generated at LO (NLO) for the analysis of the 8\TeV data use the CTEQ6L~\cite{Pumplin:2002vw}
(CTEQ6M~\cite{PhysRevD.82.074024}) set of parton distribution functions (PDFs), while those intended for the analysis of the 13\TeV data use the NNPDF3.0~\cite{Ball:2014uwa} PDF set.
The parton-level samples are interfaced to \PYTHIA 6.426 for the 8\TeV data, and to \PYTHIA 8.205~\cite{Sjostrand:2007gs} for the 13\TeV data for parton showering and hadronization, with the Z2$^\ast$~\cite{Chatrchyan:2013gfi,Khachatryan:2015pea} and CUETP8M1~\cite{Khachatryan:2015pea} tune parameter sets used, respectively, for the underlying event activity.
The cross sections and branching
fractions recommended by the LHC Higgs cross section working group for
center-of-mass energies of 8 and 13\TeV~\cite{deFlorian:2016spz}
are assumed.
After the generation step, the events are
processed by the full CMS detector simulation with \GEANTfour~\cite{GEANT}. Multiple
$\Pp\Pp$ interactions in each bunch crossing in each recorded event
(pileup) are simulated.
These
events are then weighted to reproduce the distribution of
the number of interactions observed in data in 2012\,(2016) for the 8\,(13)\TeV data, the average values of which were 21 and 23 interactions, respectively.
The
trigger efficiencies measured using the method described
above are applied to the simulated SM Higgs boson events as a correction, and the associated statistical and systematic uncertainties are propagated to the expected signal yields.

Events corresponding to the SM background processes mentioned in Section~\ref{sec:intro} are simulated using various generators. The diphoton background is modeled with the
\SHERPA 1.4.2\,(2.2.0)
\cite{Gleisberg:2008ta} generator for the analysis of the 8\,(13)\TeV data; it includes the Born processes
with up to 2\,(3) additional jets,
as well as the box processes at
LO.  Multijet and \GAMJET\ backgrounds are modeled with \PYTHIA 6.426\,(8.205) in the case of the 8\,(13)\TeV data,
with a
filter~\cite{HIG-13-001,CMS-PAS-HIG-16-040} applied at generator level in order to
enhance the production of jets with a large fraction of electromagnetic
energy.
Drell--Yan events are simulated at LO with
\textsc{MadGraph5} 1.3.30~\cite{Alwall:2014hca} and at NLO with \POWHEG 1.0~\cite{powheg-Zjj} in the case of the 8\TeV data, and entirely at NLO with
\MGvATNLO 2.2.2 for the 13\TeV data. All background events are generated using the same PDF sets and simulated under the same conditions as the SM Higgs boson events described above.
The background events are used in the calculation of energy scale and smearing corrections, preselection and photon identification efficiencies, training of the multivariate boosted decision trees (BDTs) used in the analysis, estimations of systematic
uncertainties, and for validation. In particular, the Drell--Yan events are used to obtain initial values for some of the parameters used to model the shape of the small background contribution from dielectron decays of the $\PZ$ boson, which can be misidentified as photon pairs.  As in~\cite{HIG-13-001} and~\cite{CMS-PAS-HIG-16-040}, the background estimation is extracted
from data.

\subsection{Photon reconstruction, event selection and classification}
\label{sec:selection}
\newcommand{\ptga}{\ensuremath{p_\mathrm{T}^{\gamma1}}\xspace}
\newcommand{\ptgb}{\ensuremath{p_\mathrm{T}^{\gamma2}}\xspace}
The same diphoton vertex identification is used as in~\cite{HIG-13-001} (~\cite{CMS-PAS-HIG-16-040}) for the 8\,(13)\TeV data.
For both data sets, a BDT is used to select a diphoton vertex from the set of all reconstructed primary vertices, incorporating as input variables the sum of the squared transverse momenta of the charged particle tracks associated with the vertex, and two variables that quantify the vector and scalar balance of $\pt$ between the diphoton system and the charged particle tracks associated with the vertex. Furthermore, if either photon is associated with any charged particle tracks that have been identified as resulting from conversion, the pull between the longitudinal positions of the primary vertex obtained from the conversion tracks alone and from all associated tracks is added to the BDT input variable set, and, in the case of the 13\TeV data, the number of conversions.

The same photon reconstruction
is used as in~\cite{HIG-13-001} (~\cite{CMS-PAS-HIG-16-040}) for the 8\,(13)\TeV data.   For the 8\TeV data, photon candidates are reconstructed from energy deposits in the ECAL grouped into extended clusters or groups of clusters known as ``superclusters". In the EB, superclusters are formed from five-crystal-wide strips in $\eta$, centered on the locally most energetic crystal, and have a variable extension in $\phi$. In the EE detectors, where the crystals are arranged according to an $x$--$y$ rather than an $\eta$--$\phi$ geometry, matrices of 5$\times$5 crystals, which may partially overlap and are centered on a locally most energetic crystal, are summed if they lie within a narrow $\phi$ road.  For the 13\TeV data, photon candidates are reconstructed as part of the global event reconstruction, as described in~\cite{CMS-PRF-14-001}.
 First, cluster ``seeds" are identified as local energy maxima above a given threshold. Second, clusters are grown from the seeds by aggregating crystals with at least one side in common with a clustered crystal and with an energy in excess of a given threshold. This threshold represents approximately two standard deviations of the electronic noise in the ECAL, and amounts to 80\MeV in the EB and, depending on $\abs{\eta}$, up to 300\MeV in the EE detectors. The energy of each crystal can be shared among adjacent clusters assuming a Gaussian transverse profile of the electromagnetic shower. Finally, clusters are merged into superclusters.

For both data sets, the energy of photons is computed from the sum of the energy of the clustered crystals, calibrated and corrected for changes in the response over time~\cite{Chatrchyan:2013dga}. The preshower energy is added to that of the superclusters in the region covered by this detector. To optimize the resolution, the photon energy is corrected for the containment of the electromagnetic shower in the superclusters and the energy losses from converted photons~\cite{CMS:EGM-14-001}. The correction is computed with a multivariate regression technique that estimates simultaneously the energy of the photon and its uncertainty. This regression is trained on simulated photons using as the target the ratio of the true photon energy and the sum of the energy of the clustered crystals. The inputs are shower shapes and position variables---both sensitive to shower containment and possible unclustered energy---preshower information, and global event observables sensitive to pileup.

Photon candidates are subject to
a preselection
that imposes requirements
on
$\pt$, hadronic leakage, and shower shape, and that
uses an
electron veto
to reject photon candidates geometrically matched to a hit in the pixel detector.
The preselection is designed to be slightly more stringent than the trigger requirements.
A
photon identification BDT combining lateral shower shape variables, isolation variables, the median energy density, the pseudorapidity, and the raw energy is used to separate prompt photons from nonprompt photons
resulting from neutral meson decays~\cite{HIG-13-001,CMS-PAS-HIG-16-040}.  Each photon candidate must satisfy
the preselection
requirements as well as a
requirement on the minimum value of the photon identification BDT output.  As in~\cite{HIG-13-001,CMS-PAS-HIG-16-040},
the efficiencies of the minimum photon identification BDT output requirement and preselection criteria (except for the electron
veto requirement) 
are measured with a tag-and-probe
technique using $\PZ\to\EE$ events. The
fraction of photons that satisfy the electron veto requirement is measured with
$\PZ\to\MM\gamma$ events, in which the photon is produced by final-state
radiation
providing a sample of prompt photons with purity higher than 99\%. The ratios of the efficiencies in data and simulation
are used to correct the signal efficiency in simulated signal samples and the associated statistical and systematic uncertainties are propagated to the expected signal yields.

The analysis uses all events
that contain a diphoton pair where each of the photons in the pair
satisfy a requirement
on the ratio of its
$\pt$ value to the invariant mass of the diphoton system, $\mgg$.  Specifically, in the case of the 8\,(13)\TeV data, the requirements are
$\ptga/\mgg>28.0/80.0=0.35$ $(30.6/65.0=0.47)$ and $\ptgb/\mgg>20.0/80.0=0.25$ $(18.2/65.0=0.28)$.
Here, $\gamma 1$ ($\gamma 2$) refers to the photon candidate with the highest (next-highest)
$\pt$ value.
The use of $\pt$ thresholds scaled by $\mgg$~\cite{HIG-13-001,CMS-PAS-HIG-16-040} is intended to prevent
a distortion of the low end of the
diphoton mass spectrum that results if a fixed threshold is used;
in particular, the minimum $\pt$ values in the above fractions, 28\,(30.6)\GeV and 20\,(18.2)\GeV for the 8\,(13)\TeV data,
are chosen to be slightly higher than those of the HLT paths, \ie, 26\,(30)\GeV and 18\GeV for the 8\,(13)\TeV data,
to further guard against
distortion of the spectrum.  Finally, the diphoton system invariant mass must lie within
the range
65\,$(75)<\mgg<120\GeV$ in the case of the 13\,(8)\TeV data.

A multivariate event classifier~\cite{HIG-13-001,CMS-PAS-HIG-16-040}
is
used to discriminate between diphoton events from Higgs boson decays
and those from the diphoton continuum, to further reduce background from events containing jets misidentified as isolated photons, and to assign a high score to events with good diphoton mass resolution.
It incorporates the kinematic properties of the diphoton system (excluding
$\mgg$), a per-event estimate of the diphoton
mass resolution, and the photon identification BDT output values.
The events are separated into classes based on the classifier score, with a minimum score below which they are rejected.  The number of classes and their boundaries
are determined so as to maximize the expected signal significance.
Four (three) classes are used
for the 8\,(13)\TeV data; they are referred to as 0, 1, 2, and
3\,(0, 1, and 2), where class 0 contains the events with greatest expected sensitivity.
The fraction of events containing more than one diphoton candidate is of order $10^{-4}$. In these cases, the candidate assigned to the highest sensitivity class is selected; should this class still contain multiple diphoton candidates, the candidate with the highest value of $\ptga+\ptgb$ is then selected.
\section{Signal parametrization}
\label{sec:signal}
In order to perform a statistical interpretation of the
data, it is necessary to have a description of the signal
that
includes the overall product of the efficiency and acceptance,
as well as the shape of the diphoton mass distribution in each of the
event classes. The simulated SM Higgs boson events are used
to
construct a parameterized signal model
that is defined continuously for any value of
Higgs boson mass between 80\,(70) and 110\GeV, for the
8\,(13)\TeV data.  The photon energy resolution predicted by the simulation is modified by a Gaussian smearing determined from the comparison between the $\PZ\to\EE$ line-shape in data and simulation, where the electron energies have been corrected with factors developed for photons, using the same
procedure as that described in~\cite{HIG-13-001,CMS-PAS-HIG-16-040}.  The amount of smearing is extracted differentially in  bins of $\abs{\eta}$ and
the $\RNINE$ shower shape variable~\cite{CMS:EGM-14-001},
defined as the energy sum of 3$\times$3 crystals centered on the most energetic crystal in the ECAL cluster divided by the energy of the cluster.  The trigger and preselection efficiency corrections described in Sections~\ref{sec:triggerSim} and~\ref{sec:selection}, respectively,
are also applied to the simulated signal events.

Since the shape of the \mgg distribution changes considerably depending on whether the vertex associated with the candidate diphoton is correctly identified, separate fits are made to the distributions for the correct and incorrect primary vertex selections when constructing the signal model.  Events are considered to have the correct primary vertex if the vertex associated with the candidate diphoton is within 1\unit{cm} of the true vertex.  For these events the signal shape
is dominated by
ECAL response and reconstruction, and
is modeled empirically
by a sum of
between three and six
(three
and four)
Gaussian functions in the case of the 8\,(13)\TeV data, depending on the event class. The signal shape for events with an incorrect primary vertex selection is smeared significantly by the variation in the $z$-coordinate position of the selected primary vertex with respect to the true Higgs boson production
vertex.  The signal shape for these events is modeled
by a sum of
between one and four
(two
and three) Gaussian functions in the case of the 8\,(13)\TeV data, depending on the event class. In
both cases, the means, widths, and relative fractions of the Gaussian functions are
determined by the fits.

\begin{figure*}[htb]
 \centering
   \includegraphics[width=0.49\textwidth]{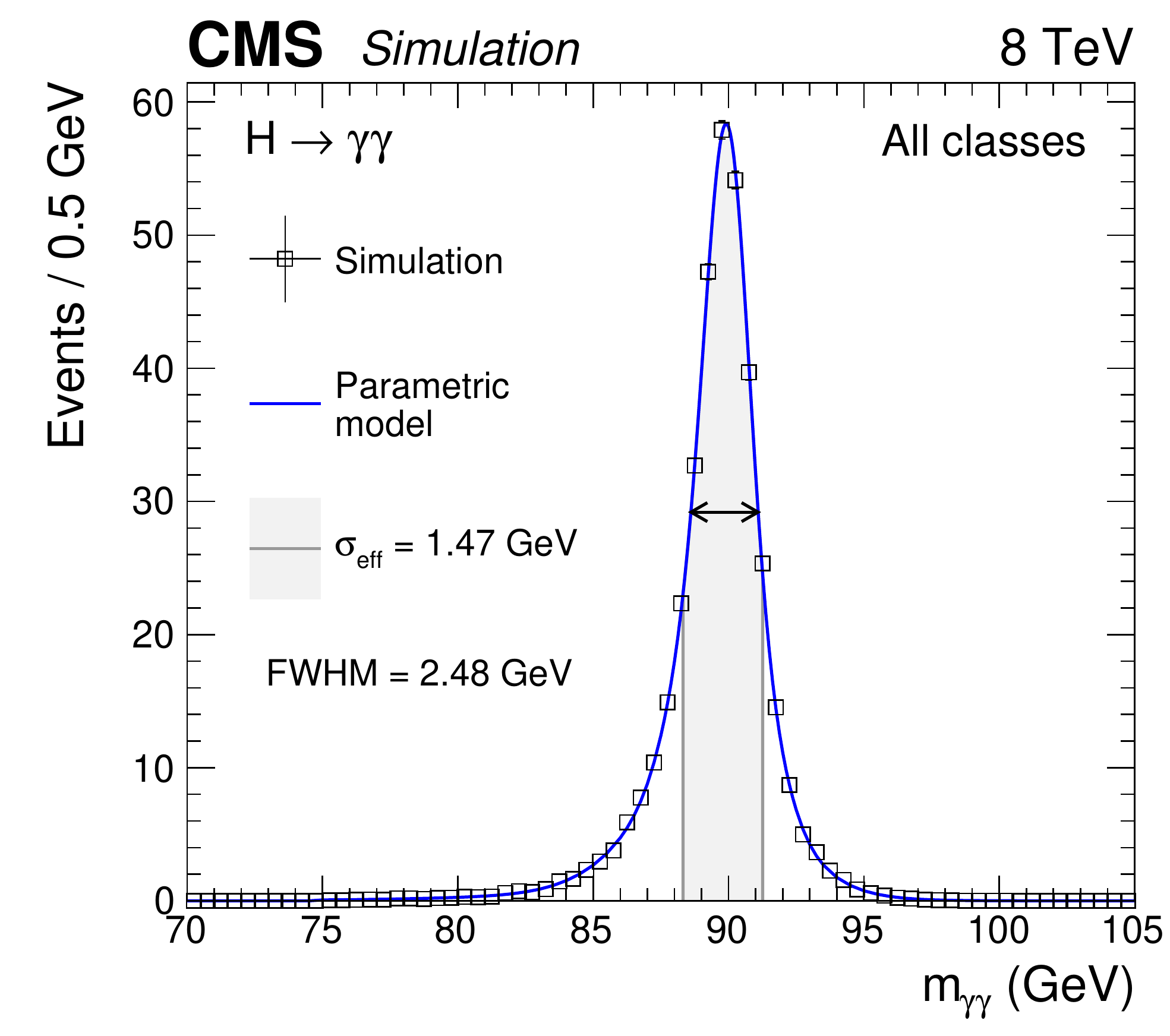}
   \includegraphics[width=0.49\textwidth]{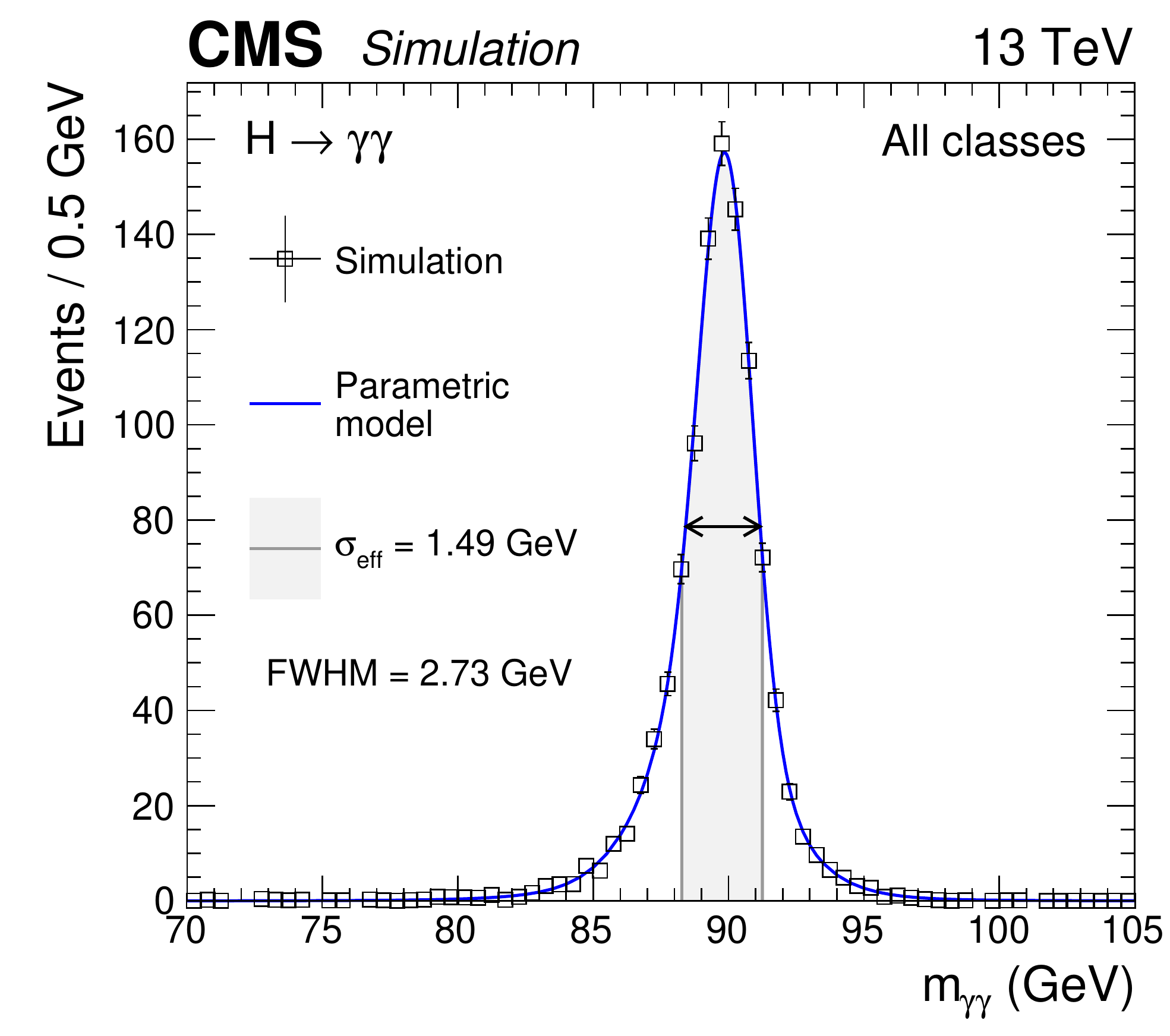}
\caption{
      \label{sigPlotsComb}
        Full parameterized signal shape, integrated over all
event classes, in simulated signal events with $\mH=90\GeV$ at $\sqrt{s}=8\TeV$ (left) and
13\TeV (right). The
open points are the weighted MC events and the blue lines
the corresponding parametric models. Also shown are the  $\sigma_{\text{eff}}$ values and the shaded region limited by $\pm\sigma_{\text{eff}}$, along with the
FWHM values,
indicated by the position of the arrows on each distribution.
            }
\end{figure*}

The full signal model for all values of $\mH$ is
obtained by
linear interpolation of each of the fitted parameters.
The final parameterized shapes
for the combination of all production mechanisms, for all event classes, weighted by their SM cross sections
are shown in Fig.~\ref{sigPlotsComb} for a Higgs boson mass of
90\GeV for the 8
and 13\TeV data.
Also shown are the full width at half maximum (FWHM) value and
the value of the effective
standard deviation for signal ($\sigma_{\text{eff}}$), which is defined as half the width of the narrowest interval containing 68.3\% of the
invariant mass distribution.
The product of efficiency and acceptance of the signal model ranges from 36.2\,(22.7)\% for $\mH=80$ (70)\GeV to 40.4\,(26.5)\% for $\mH=110$ (110)\GeV in the case of the 8\,(13)\TeV data.

\section{Background estimation}
In this analysis, as in~\cite{Chatrchyan:2012xdj,HIG-13-001,CMS-PAS-HIG-16-040},
the background is modeled by fitting analytic functions to the observed diphoton mass distributions, in each of the event classes.
The fits are performed over the range 75\,$(65) < \mgg < 120\GeV$ for the 8\,(13)\TeV data. In the case of the 8\TeV data, a single fit function is chosen for each class after a
study of the potential bias
in the estimated background, which is required to be negligible, following the method used in~\cite{Chatrchyan:2012xdj}. For the 13\TeV data, as in~\cite{HIG-13-001,CMS-PAS-HIG-16-040},
the model is
determined from data with the discrete profiling method~\cite{DiscreteProfilingMethod}, which treats the choice of the background function as a discrete parameter in the likelihood fit to the data and estimates the systematic uncertainty associated with the choice of a particular function.

Since the search mass range of this analysis includes the
\PZ boson peak region, a significant potential background source is
Drell--Yan dielectron production that, through misidentification, could appear to result in two isolated photons.
Therefore, an explicit component intended to describe the background from the Drell--Yan process in which the two apparent isolated photons 
survive all the selection requirements as stated in Section~\ref{sec:selection},
is added to the smoothly falling polynomial
distribution
used to model the background
in~\cite{Chatrchyan:2012xdj, HIG-13-001, CMS-PAS-HIG-16-040}.
This additional component, referred to as ``doubly misidentified" events, is modeled
with
a double-sided Crystal Ball (DCB) function, which is a  modification of the Crystal Ball function~\cite{CrystalBall}  with an exponential tail on
both sides.
The DCB function
is characterized by seven parameters:
the number of events
for normalization, the Gaussian mean and standard deviation, and the
four additional shape parameters
$\alpha_{\text{L}}$, $n_{\text{L}}$, $\alpha_{\text{R}}$, and $n_{\text{R}}$, where
$\alpha_{\text{L,R}}$ and $n_{\text{L,R}}$ refer, respectively, to the slope and normalization of the left-hand (L) and right-hand (R) exponential tails.
The values of the DCB
shape parameters are
determined by fitting the diphoton invariant mass distribution in
a sample of
simulated Drell--Yan
doubly misidentified events for each event class.
Because of the
small size of the simulated event
sample, we fix two of the six DCB
shape parameters, $\alpha_{\text{L}}$ and $\alpha_{\text{R}}$,
to make the fit more stable. The fixed values are different in each event class
and are obtained
using the normalized $\chi^{2}$ value for the 8\TeV data, and the minimal maximum pull
value for the 13\TeV data,
as a figure of merit. In each class the value of the
mean, which coincides with the peak position, lies somewhat below the nominal \PZ boson mass value. This is due to
the fact that the electrons surviving the photon selection requirements (in particular the electron veto) have in general been poorly reconstructed, for example having undergone wide-angle bremsstrahlung of high-energy photons;
furthermore, the electron energies have been corrected
with factors developed for photons.

For both the choice of the single fit function in the case of the 8\TeV data, and the application of the discrete profiling method
in the case of the 13\TeV data,
members of several families of analytic functions, including
exponential,
power law,
Bernstein,
and Laurent
series are considered, each summed with a DCB function.
    The
maximum order term in each series
is determined using an F-test~\cite{ftest}.  In
the analysis of the 13\TeV data,  the minimum order of the series is determined as well, using
a goodness-of-fit test.

In the analysis of the 8\TeV data these functions, called ``truth models", are used to generate
MC pseudo-data sets
that are fitted
with candidate functions
from the same
families of an order within the range determined by the above tests.  The bias for a given candidate function to fit a given truth model is defined as the average pull
of the fitted signal strength modifier over the set of relevant generated
pseudo-data sets, and is required to be less than 0.14 to be considered negligible. This amount of bias necessitates an increase in the uncertainty in the frequentist coverage
of the signal strength of less than 1\%, which is deemed acceptable.  The final background function is chosen from the candidate functions that fit all truth models in a given event class with negligible bias.

In the discrete profiling method used for the analysis of the 13\TeV data, when fitting these functions to the background \mgg distribution, the value of twice the negative logarithm of the likelihood (2NLL) is minimized.  A penalty is added to the 2NLL value to take into account the number of floating parameters, including
the fraction of background events attributed to the component arising from the doubly misidentified events (DCB fraction), in each candidate function.

In both methods, the normalization of the Drell--Yan background is
determined in the fit.
The shape parameters are constrained to the constant values
that are obtained by fitting the
doubly misidentified Drell--Yan events, as described above.   In particular, the value of the Gaussian standard deviation in each event class is greater than the corresponding value of $\sigma_{\text{eff}}$ in the signal model by a factor of up to 2.

For the analysis of the 8\TeV data, the sum of a fifth-order Bernstein polynomial and the DCB function is chosen as the final background model for
event classes 1, 2, and 3.
For class 0,
a fourth-order Bernstein polynomial is used.
For the 13\TeV data,
a third-order
exponential series plus the DCB function is chosen for classes 0 and 2, and a first-order
power-law series plus the DCB function for class 1.
The DCB fractions for these chosen models in the subset of the diphoton mass range extending
from 85
to 95\GeV, the most relevant for dielectron background from the Drell--Yan process, are,
for the 8\,(13)\TeV data,  3.0, 5.6, 2.6, and 5.1\,(3.0, 3.1, and 3.3)\%, respectively, for event classes 0, 1, 2, and 3\,(0, 1, and 2).

\begin{figure*}[!thb]\centering
   \includegraphics[width=0.49\textwidth]{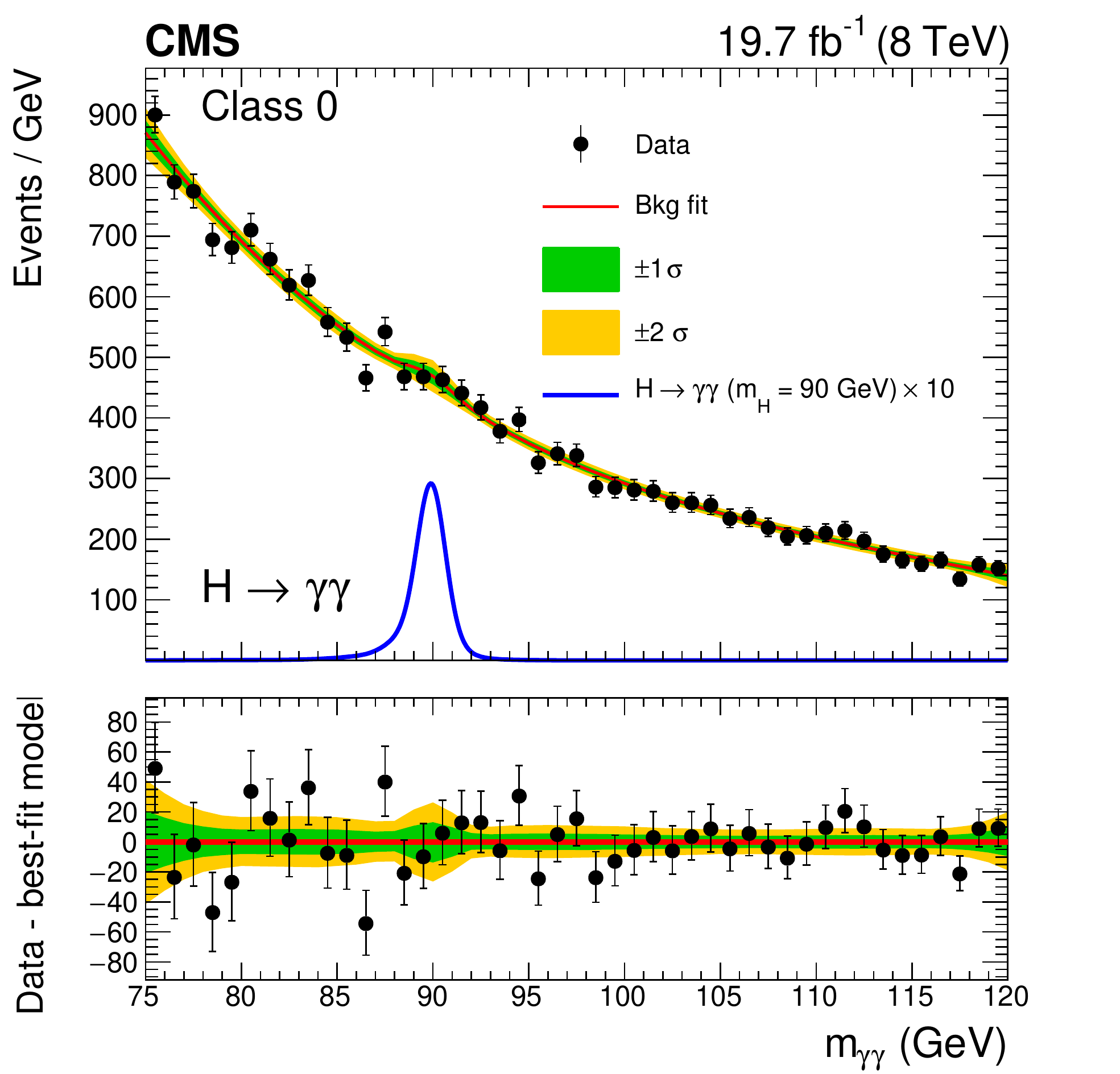}
   \includegraphics[width=0.49\textwidth]{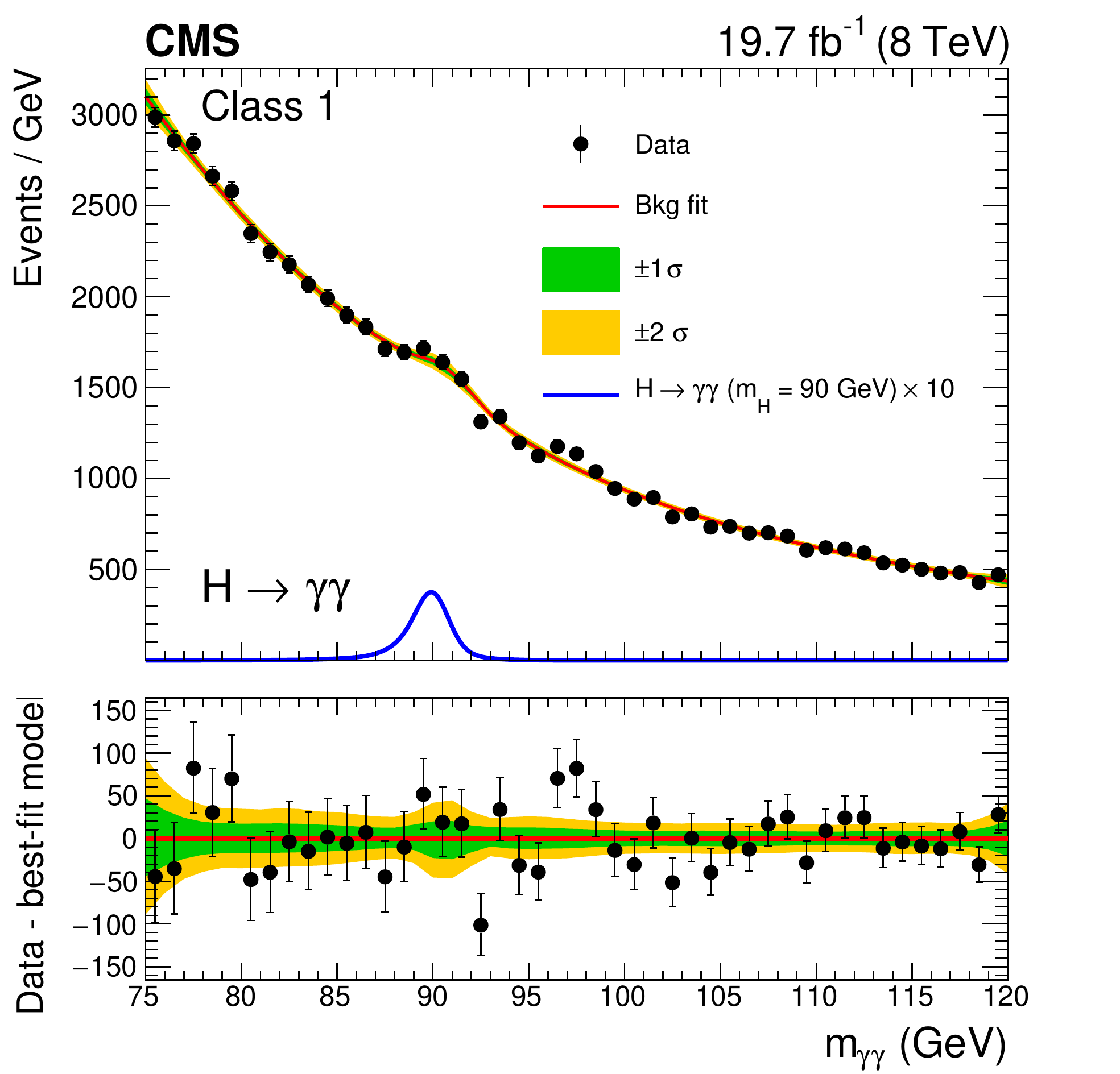} \\
   \includegraphics[width=0.49\textwidth]{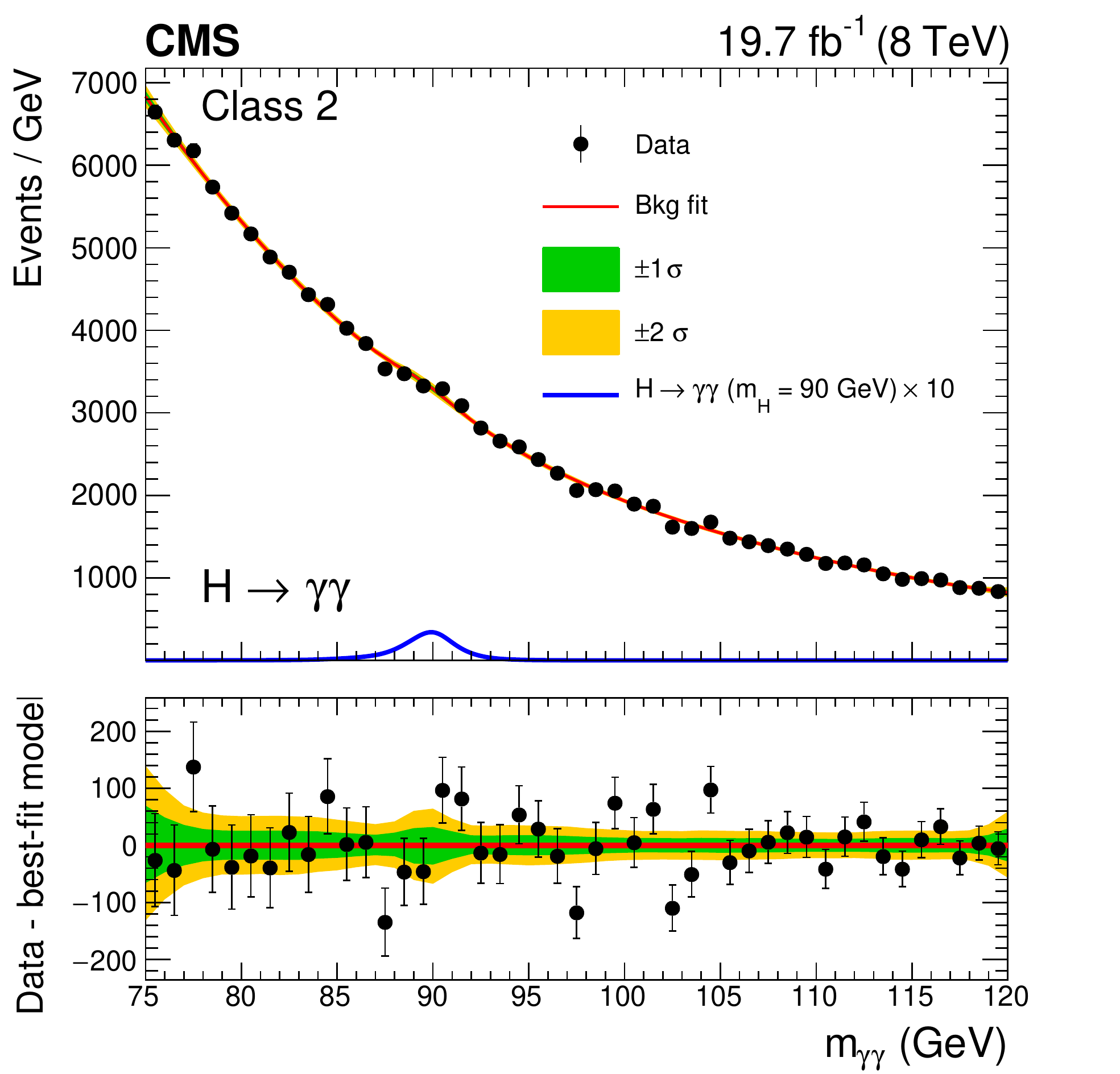}
   \includegraphics[width=0.49\textwidth]{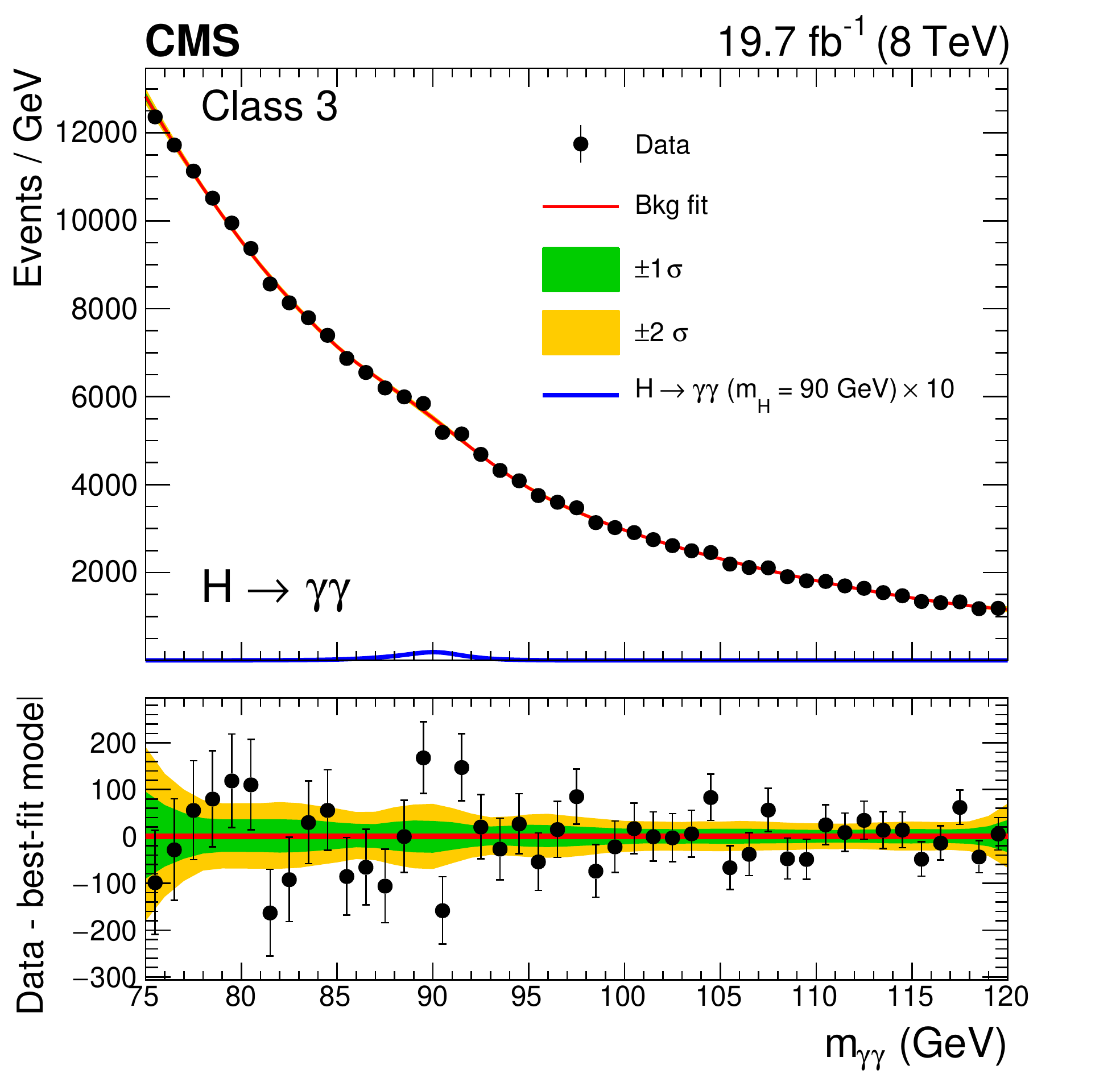}
   \caption{Background model fits using the chosen ``best-fit" parametrization
to data in the four event classes at $\sqrt{s}=8\TeV$.
The corresponding signal
   model for each class for  $\mH=90\GeV$, multiplied by 10, is also shown. The
one- and two-$\sigma$ bands
reflect the uncertainty in the background model normalization associated with the statistical uncertainties of the fits, and are shown for illustration purposes only. The difference
between the data and the best-fit
model is shown
in the lower panels.}
     \label{bkgModelFinal}
   \end{figure*}

\begin{figure*}[!htb]\centering
 \includegraphics[width=0.49\textwidth]{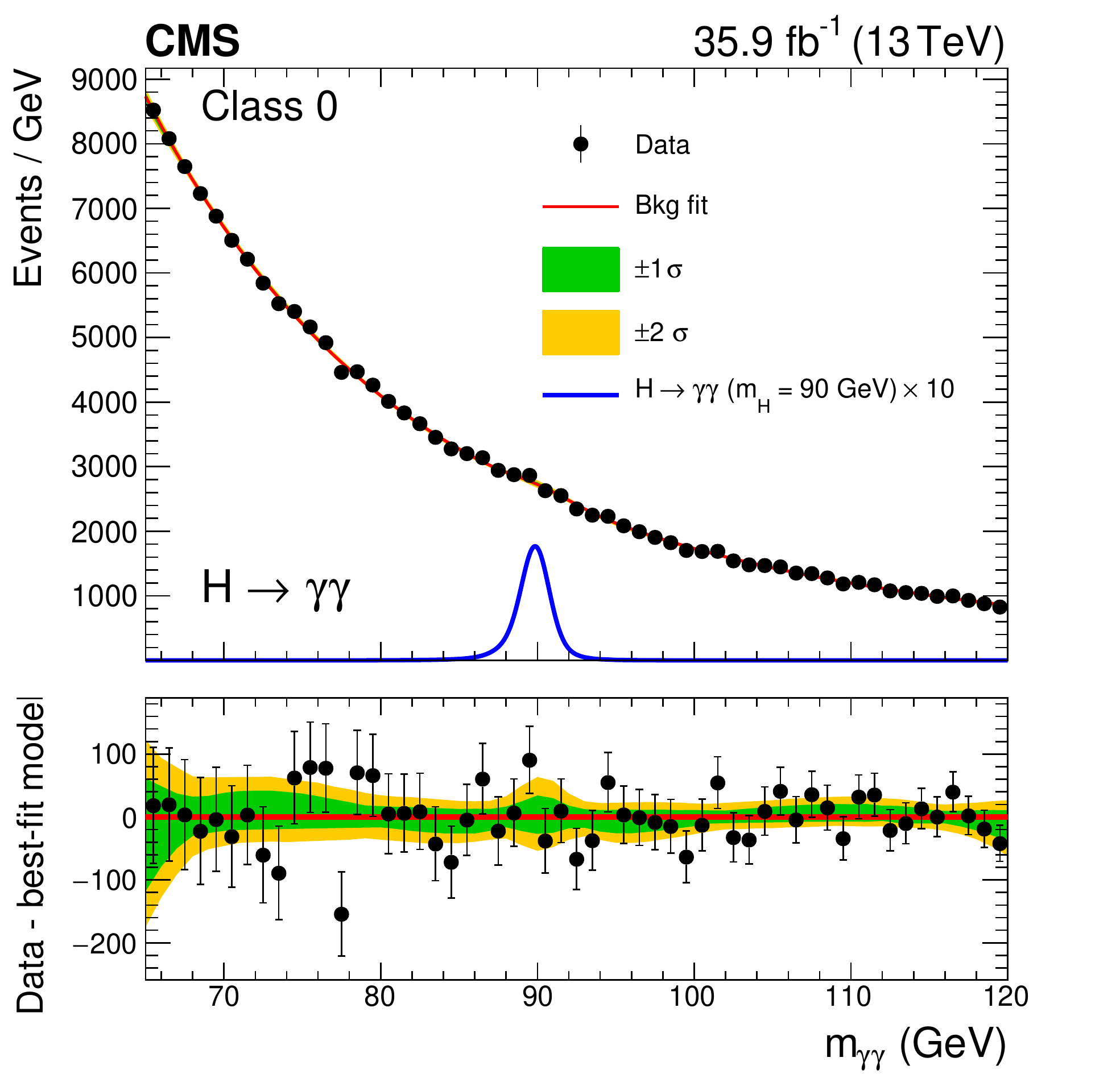}
 \includegraphics[width=0.49\textwidth]{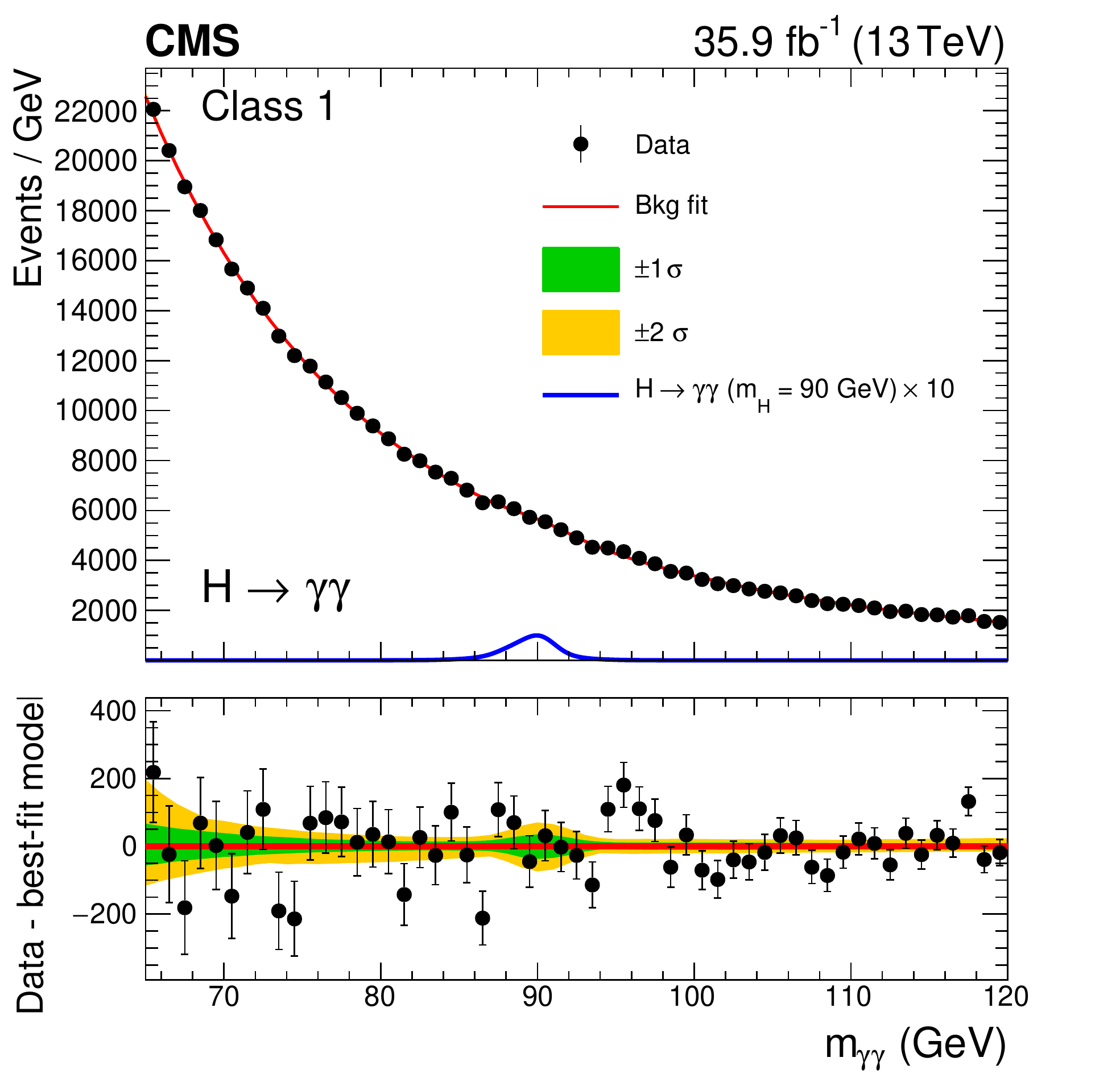} \\
 \includegraphics[width=0.49\textwidth]{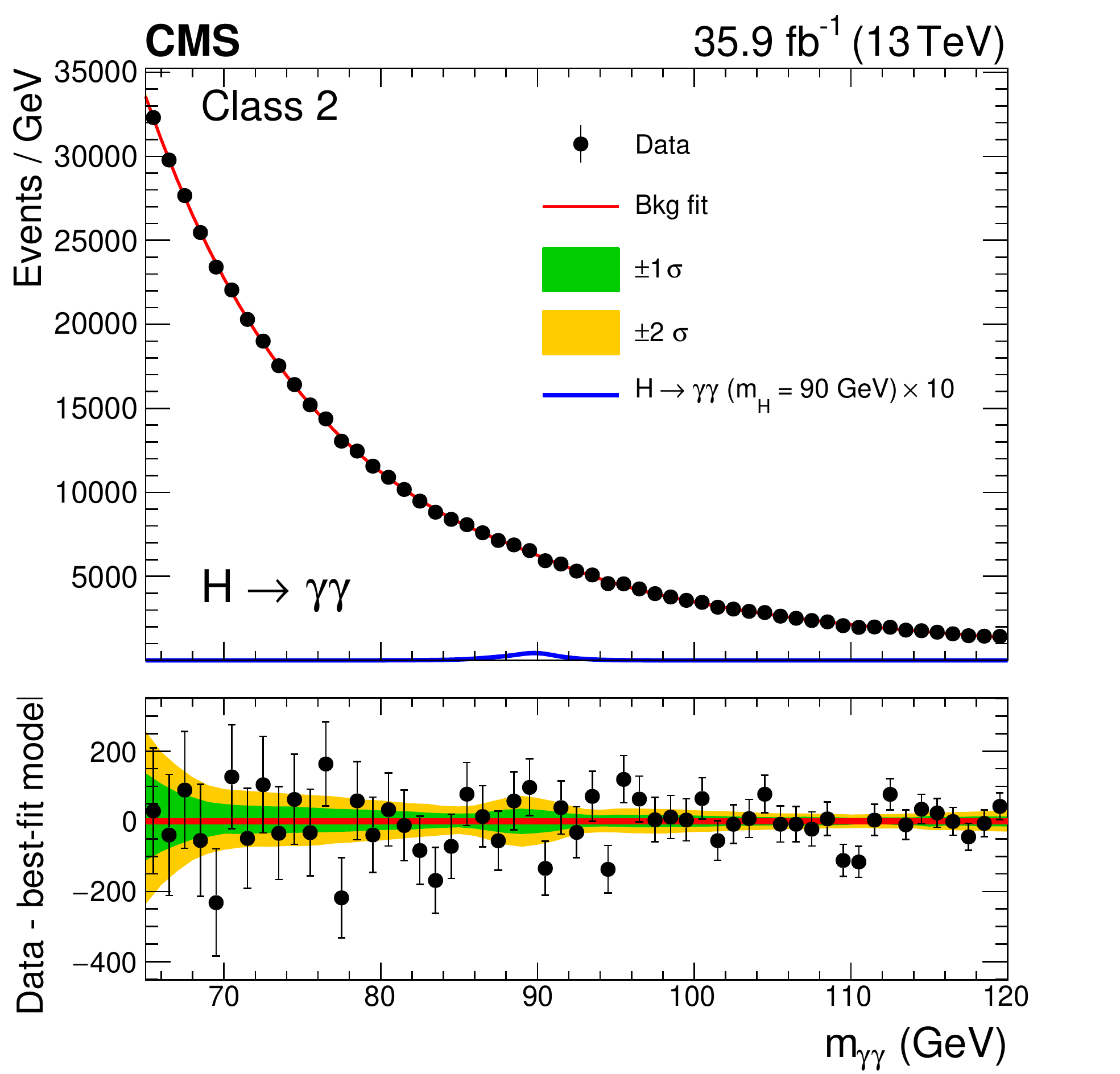}
 \caption{Background model fits using the chosen ``best-fit" parametrization
to data in the three event classes at $\sqrt{s}=13\TeV$.
The corresponding signal
   model for each class for  $\mH=90\GeV$, multiplied by 10, is also shown. The
one- and two-$\sigma$ bands
reflect the uncertainty in the background model normalization associated with the statistical uncertainties of the fits, and are shown for illustration purposes only. The difference
between the data and the best-fit
model is shown
in the lower panels.}
 \label{fig:statAnalysisBkgValidationPlotsUntagged}
\end{figure*}

Binned likelihood fits of the chosen background models to the observed diphoton mass distribution,
assuming no signal, are shown for all the event classes in Fig.~\ref{bkgModelFinal}
(\ref{fig:statAnalysisBkgValidationPlotsUntagged}) for the 8\,(13)\TeV data.  
The
one- and two-standard deviation ($\sigma$) bands
include only the uncertainty
in the background model normalization associated with the statistical uncertainties of the fits, and are thus shown for illustration purposes only. They are obtained using an extended
likelihood fit parametrized in terms of the background yield in a window
that is the size of the bin widths in Figs.~\ref{bkgModelFinal} and~\ref{fig:statAnalysisBkgValidationPlotsUntagged}.  The corresponding signal model for $\mH=90\GeV$, multiplied by 10, is also shown for illustration purposes.

\section{Systematic uncertainties}

Many of the systematic uncertainties relevant
to the analyses performed in~\cite{Chatrchyan:2012xdj,HIG-13-001,CMS-PAS-HIG-16-040}
also apply to
this analysis and are described briefly below.
Additional
uncertainties specific to this analysis are described in more detail.

\subsection{Uncertainties evaluated at the per-photon level}

The systematic uncertainties in the
shape of the photon identification BDT distribution and in the
per-photon energy resolution described in~\cite{HIG-13-001,CMS-PAS-HIG-16-040}
are
applied in this analysis.
These uncertainties
propagate to the multivariate event classifier value, giving rise to the migration of events from one class to another,
and to variations in the per-event efficiency in each class and for each production process. The uncertainties
are evaluated using a signal sample with $\mH=105$ (90)\GeV for the analysis of the 8\,(13)\TeV data.  For the 8\TeV data, the largest variation in efficiency due to the photon identification BDT distribution shape is
5.9\%,
for the VBF process in event class 3.
For the 13\TeV data the largest variation is
14.6\% for the VBF process in event class 2, with
other processes in class 2
having variations of less than
11\%,  and variations in
the other classes
being below 5\%.
The largest variation in the efficiency due to
the per-photon energy resolution
applicable to the 8\TeV data
is 13.7\%
for the
$\Pg\Pg\PH$ process in class 0; otherwise the
variations are below 9\%. For the
13\TeV data,
the largest
variation
is 7\%
for the VBF process in class 2; otherwise the
variations are below 5\%.

For the 8\,(13)\TeV data, uncertainties in the trigger efficiencies give rise to
efficiency variations of 1 (less than 1)\%,
and
in the scale factors of the preselection, of less than 1.5\,(5.5)\%.
In the case of the 13\TeV data, the uncertainties in the scale factors of the electron veto and
of the
minimum
value of the photon identification BDT are considered as supplemental sources of
efficiency variations, which amount to less than 2\% for each.

The uncertainties in the measurement and in the correction of the photon energy scale in data, and in the correction of the energy resolution in simulation, arising from the methodology exploiting $\PZ\to\EE$  events as described in Section~\ref{sec:signal} and~\cite{HIG-13-001,CMS-PAS-HIG-16-040}, are
calculated in the same bins as the corrections themselves.
Uncertainties arising from
modeling of the material budget and of nonuniformity of light collection (the fraction of crystal scintillation light detected as a function of its longitudinal depth when emitted), nonlinearity in the photon energy scale between data and simulation, imperfect electromagnetic shower simulation, and vertex finding~\cite{HIG-13-001,CMS-PAS-HIG-16-040},
are propagated to the parametric signal model, where they result in uncertainties in the diphoton efficiency, mass scale, and resolution.

\subsection{Uncertainties evaluated at the per-event level}

The per-event systematic
uncertainty in the total integrated luminosity, estimated from data
~\cite{CMS-PAS-LUM-13-001,CMS-PAS-LUM-17-001},
contributes an uncertainty of 2.6\,(2.5)\%
in the signal yield for the 8\,(13)\TeV data.

The
systematic uncertainties from the theoretical predictions
considered in this analysis are of two types.  Firstly,
the
uncertainties in the signal acceptance due to changes in particle  $\pt$ and $\eta$ values, arising from variations in the
PDF and
renormalization and factorization scales,
are calculated~\cite{HIG-13-001,CMS-PAS-HIG-16-040}
using a signal sample with $\mH=105$ (90)\GeV for the analysis of the 8\,(13)\TeV data.
The CT10~\cite{PhysRevD.82.074024} PDF set (NNPDF3.0~\cite{Ball:2014uwa} PDF set using the \textsc{MC2hessian} procedure~\cite{Carrazza2015}) is used to estimate the PDF variations in the case of the 8\,(13)\TeV data.
In the case of the 13\TeV data, the effects due to variations of the strong coupling strength, $\alpS$, are also considered, following the
PDF4LHC prescription~\cite{deFlorian:2016spz,Butterworth:2015oua}.
The uncertainty of greatest magnitude due to
PDF variations,
in the 8\TeV data,
is 2\%
for the VBF production
process in event class 0; otherwise the uncertainties are below 1\% and, in many cases, well
below 1\%.
In the 13\TeV data,
the
uncertainties
are equal to or less than
0.4\%.
The largest uncertainty due to
scale variations,
in the 8\TeV data,
is 7.5\%
for the
$\Pg\Pg\PH$ production process in event class 0; otherwise
the uncertainties are below 1\%.  In the
13\TeV data, the largest
uncertainties also occur for the
$\Pg\Pg\PH$ process, with the maximum of
3.8\% again occurring in event class 0.
The uncertainties due to variations in
$\alpS$, considered for the 13\TeV data,
are typically
below 0.5\%, with the largest
uncertainty of
0.7\% occurring
for the
VBF process in event class 2.

Secondly, the uncertainties in the production cross sections for
an SM-like Higgs boson,
at center-of-mass energies of 8 and 13\TeV,
are accounted for
following
the recommendations of the LHC Higgs cross section working group~\cite{deFlorian:2016spz}.
These uncertainties are due to PDF, $\alpS$, and scale variations.
They are used in the calculation of the expected and observed limits on the product of the
production
cross section and
branching
fraction into two photons relative to the
expected value for an SM-like Higgs boson,
and in the calculations of the expected and observed local $p$-values.
The uncertainty in the branching
fraction into two photons
is neglected.

An additional source of per-event systematic uncertainty specific to this analysis is the modeling of the $\PZ$ boson resonance component
of the background.
As explained previously, the parameters of the DCB function used to model the $\PZ$ boson resonance are obtained from
doubly misidentified events, which are simulated Drell--Yan events with all selection requirements applied including the electron veto requirement.
These parameters could be different
for data and simulation.  To estimate these differences,
 we study simulated events from the Drell--Yan, diphoton,
$\GAMJET$, and QCD physics processes where one photon candidate survives all selection requirements including the electron veto, and the other survives all selection requirements but fails the electron veto
("singly misidentified" events).  We fit the invariant diphoton mass of these events in data, in
simulation including the sum of all background processes,
and in simulated Drell--Yan
events alone,
with a DCB plus an
exponential component
that describes the additional continuum background inherent in singly misidentified events. We consider the pairwise differences
among the DCB
mean and standard deviation parameters extracted from these three types of fits for each event class.
The differences are considered
statistically significant if greater than the
quadratic sum of the statistical
uncertainties from the fit.
These differences will contribute to the total systematic
uncertainty in the
DCB parameter values.
The nominal parameter values are obtained from
doubly misidentified events so
the
differences contributing to the parameter uncertainties that are estimated from singly misidentified events are doubled, to reflect the
more conservative case where the parameters of the two
photon candidates in a
doubly misidentified event
are completely correlated.

The total systematic
uncertainty in each event class
for
the
mean and standard deviation parameters,
is then the quadratic sum of:
the statistical
uncertainty from the fit
to the
doubly misidentified simulated Drell--Yan events;
the
doubled difference
between the  parameter values
from data and from the sum of all simulated background processes; and the doubled difference between the parameter values from
the sum of all simulated background processes
and from simulated Drell--Yan events alone,
determined from the singly misidentified events.  As a conservative measure in the case of the 8\TeV data, the doubled differences in the parameter values for the event class where the values are maximal are used for all four classes.

Finally, the analysis takes into account the statistical uncertainties in the values of the DCB $n_{\text{L}}$ and $n_{\text{R}}$ parameters obtained from the
fits to the doubly misidentified simulated $\PZ\to\EE$ events.
\section{Results}

Table~\ref{tab:statAnalysisSigBkgYields} shows the expected number of signal events corresponding to the
production of a hypothetical
additional SM-like
Higgs boson with $\mH=90\GeV$, from the
analyses of the 8 and 13\TeV data.  The total number is broken down
into the
contributions
from all the production processes
in each of the event classes, where the $\mathrm{V}\PH$ processes corresponding to
$\PW$
and $\PZ$ are listed separately.  Also shown are the $\sigma_{\text{eff}}$ and $\sigma_{\text{HM}}$ (defined as the FWHM divided by 2.35) values, as well as the
number of background events per \GeV estimated from the background-only fit to the data, that includes the number, shown separately, 
from the Drell--Yan process,
in the corresponding $\sigma_{\text{eff}}$ window centered on $\mH=90\GeV$, using the chosen background function.
\begin{table*}[!tb]
\centering
\topcaption{The expected number of SM-like Higgs boson signal events ($\mH=90\GeV$) per event class and the corresponding percentage breakdown per production process, for the 8 and 13\TeV data. The values of $\sigma_{\text{eff}}$ and $\sigma_{\text{HM}}$
are also shown,
along with the
number of background events (``Bkg.'') per \GeV estimated from the background-only fit to the data, that includes the number, shown separately, 
from the Drell--Yan process (``DY Bkg.''), in a $\sigma_{\text{eff}}$ window centered on $\mH=90\GeV$.
}
\cmsTable{
\begin{tabular}{lrrrrrrrccrr}
\hline
\multicolumn{2}{l}{\multirow{3}{*}{Event classes}} & \multicolumn{8}{c}{Expected SM-like Higgs boson signal yield ($\mH=90\GeV$)} & \multicolumn{1}{c}{Bkg.} & \multicolumn{1}{c}{DY Bkg.}  \\
\cline{3-10}
\multicolumn{2}{c}{} & Total & $\Pg\Pg\PH$ & VBF & $\PW\PH$ & $\PZ\PH$
& $\ttbar\PH$ & $\sigma_\text{eff}$ & $\sigma_\mathrm{HM}$ & \multicolumn{1}{c}{(${\GeVns}^{-1}$)} & \multicolumn{1}{c}{(${\GeVns}^{-1}$)}  \\
\multicolumn{2}{c}{} & & (\%) & (\%) & (\%) & (\%) & (\%) & (\GeVns) & (\GeVns) &  & \\
\hline
8\TeV & 0 &    64  &  68.9  &  14.9  &  8.8  &  4.8  &  2.5 & 0.94 & 0.78 & 467 & 30 \\
19.7\fbinv & 1 &    100  &  87.5  &  5.3  &  4.3  &  2.3  &  0.7 & 1.20 & 0.96 & 1639 & 157 \\
& 2 &    121  &  90.0  &  3.9  &  3.7  &  2.0  &  0.5 & 1.61 & 1.26 & 3278 & 145\\
& 3 &    89  &  92.2  &  2.8  &  3.0  &  1.6  &  0.3 & 2.11 & 1.68 & 5508 & 383 \\
&Total &    374  &  86.2  &  5.9  &  4.6  &  2.4  &  0.8 & 1.47 & 1.05 & 10\,892  & 715 \\
[\cmsTabSkip]
13\TeV &0 &    457  &  80.2  &  9.7  &  4.9  &  2.8  &  2.5 & 1.11 & 0.96 & 2720 & 132\\
35.9\fbinv & 1 &    395  &  90.1  &  4.1  &  3.2  &  1.7  &  0.9 & 1.69 & 1.45 & 5636 & 282\\
& 2 &    214  &  92.0  &  3.3  &  2.6  &  1.4  &  0.7 & 2.18 & 1.73 & 6256 & 274 \\
&Total &    1066  &  86.2  &  6.3  &  3.8  &  2.1 &  1.6 & 1.49 & 1.16 & 14\,612 & 688\\
\hline
\end{tabular}
}

\label{tab:statAnalysisSigBkgYields}
\end{table*}

\begin{figure}[!htbp]\centering
   \includegraphics[width=0.49\textwidth]{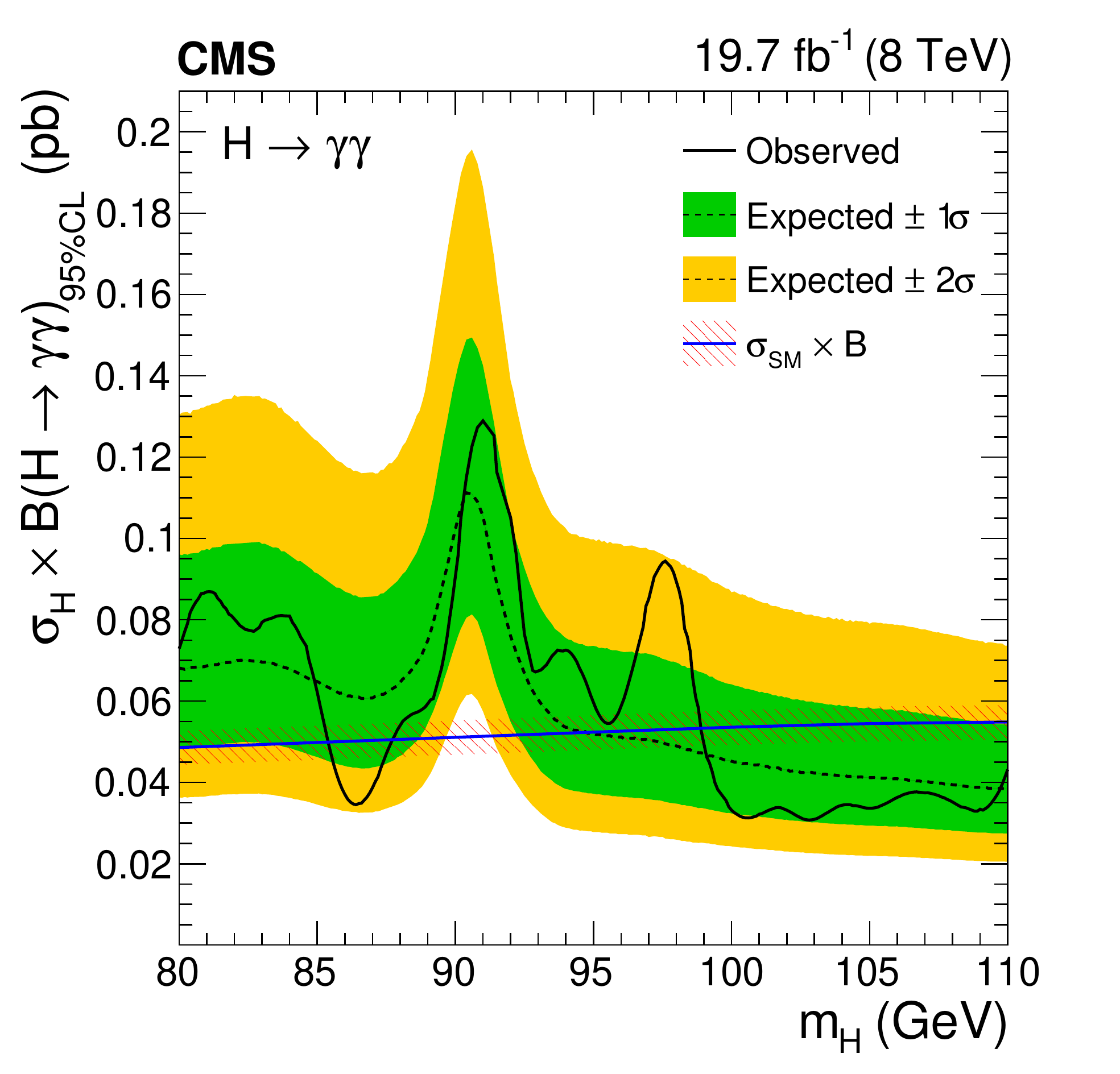}
    \includegraphics[width=0.49\textwidth]{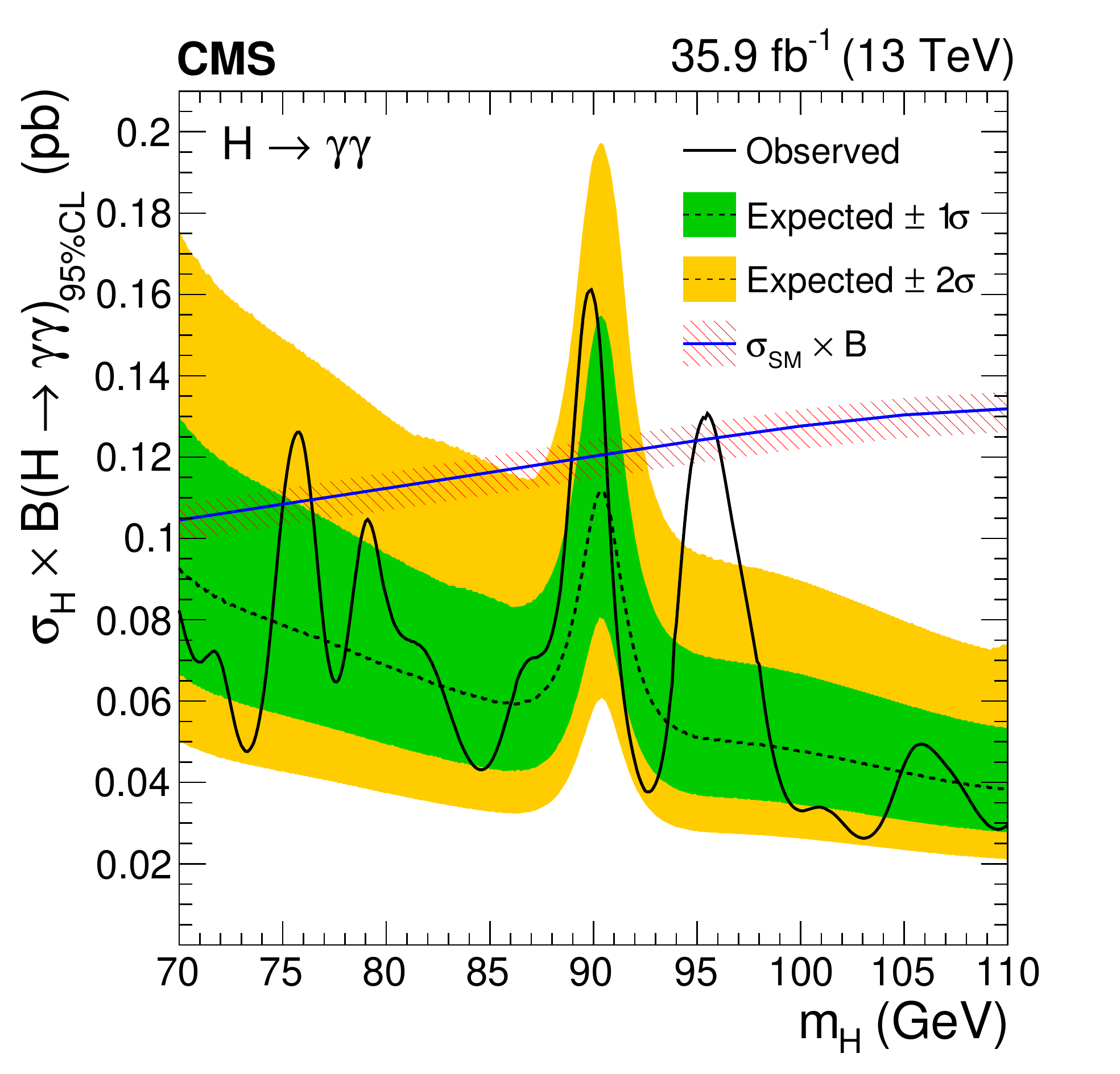}
   \caption{
      \label{8TeVLimitAbs}
       Expected and observed exclusion limits (95\% \CL, in the asymptotic approximation) on the
product of the production cross section
and branching
fraction into two photons for an
additional SM-like
Higgs boson,
from the analysis of the 8 (\cmsLeft) and 13 (\cmsRight)\TeV data. The inner
and
outer
bands indicate the regions containing the distribution of limits located within $\pm$1 and 2$\sigma$,
respectively, of the
expectation under the background-only hypothesis. The corresponding theoretical prediction for the product of the cross
section and branching fraction into two photons for an additional SM-like Higgs boson is shown as a
solid line with
a hatched band, indicating its uncertainty~\cite{deFlorian:2016spz}.
    }
\end{figure}

A simultaneous binned maximum likelihood fit to the diphoton invariant mass distributions in all event classes,
with a step size of 0.1\GeV,
 is performed over the range 75\,$(65)<\mgg<120\GeV$ for the 8\,(13)\TeV data, using
an asymptotic
approach~\cite{LHC-HCG,Junk:1999kv,Read:2000ru} with a
test statistic based on the profile likelihood ratio~\cite{Cowan:2010js}.
The expected and observed 95\% confidence level (\CL) upper limits on the product of the
cross section ($\sigma_{\PH}$)
and branching
fraction ($\cal{B}$) into two photons for an
additional SM-like
Higgs boson, from the analysis of each of the 8 and 13\TeV data sets, are presented in Fig.~\ref{8TeVLimitAbs} for the parametric
signal model.
No significant (${>}3\sigma$) excess
with respect to the expected number of background events
is observed. For the 8\TeV data, the minimum (maximum) observed upper limit on the product of the production cross section
and branching
fraction
is approximately 31\,(129)\unit{fb}, corresponding to a
mass hypothesis of 102.8\,(91.0)\GeV.
For the 13\TeV data, the minimum (maximum) observed upper limits are 26\,(161)\unit{fb}, corresponding to a
mass hypothesis of 103.0\,(89.8)\GeV.

\begin{figure*}[!htb]\centering
\includegraphics[width=0.49\textwidth]{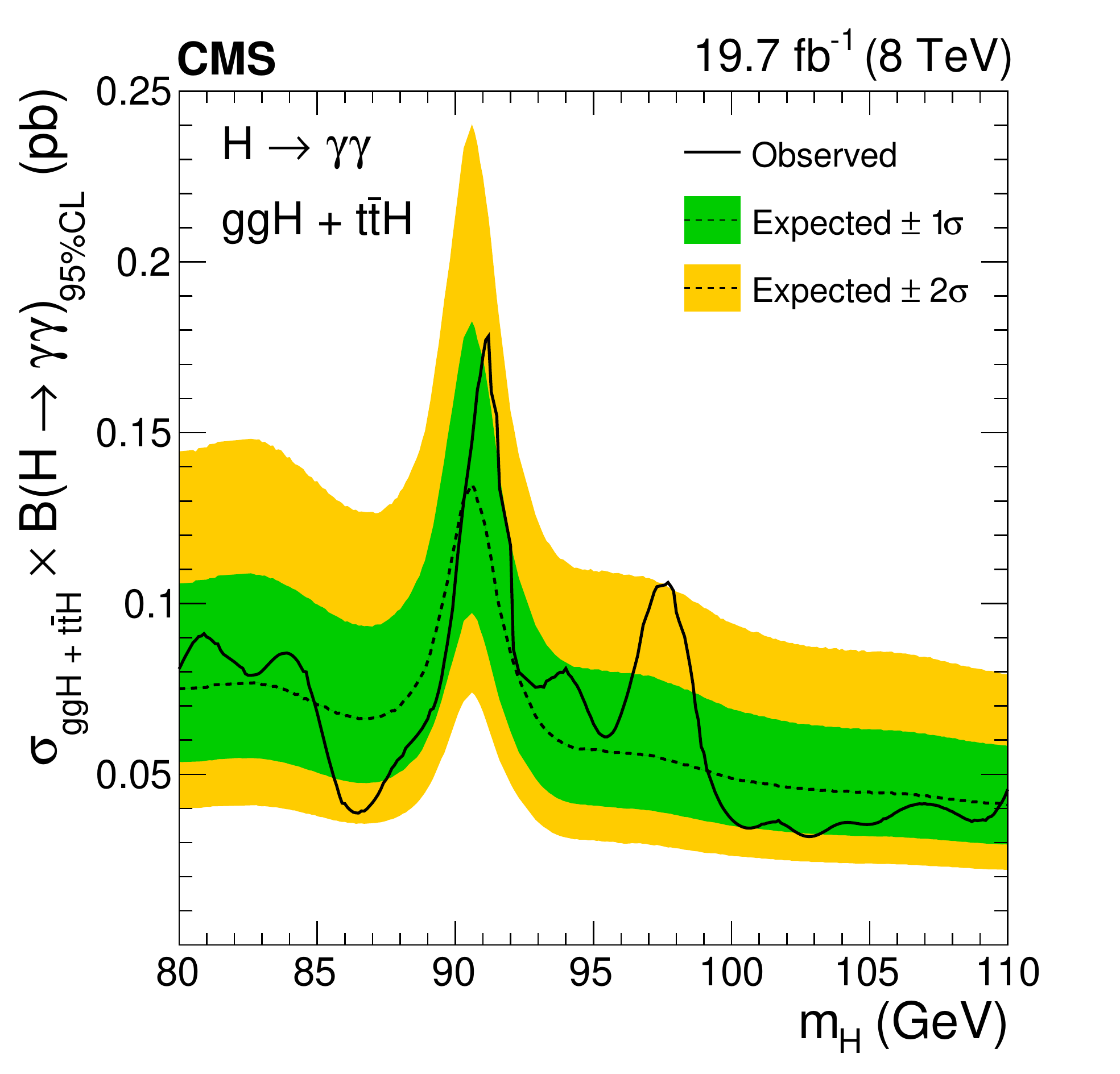}
\includegraphics[width=0.49\textwidth]{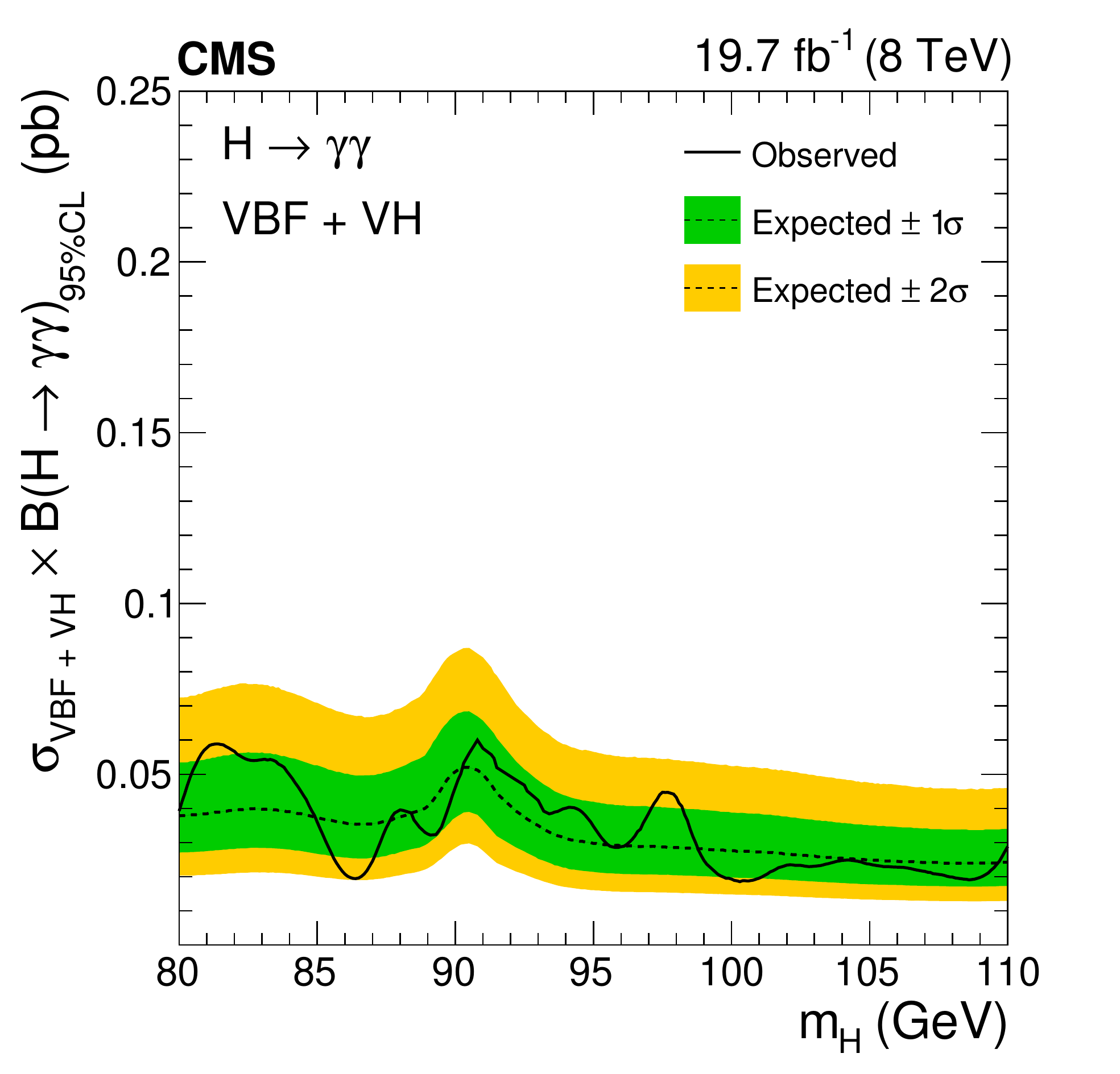} \\
\includegraphics[width=0.49\textwidth]{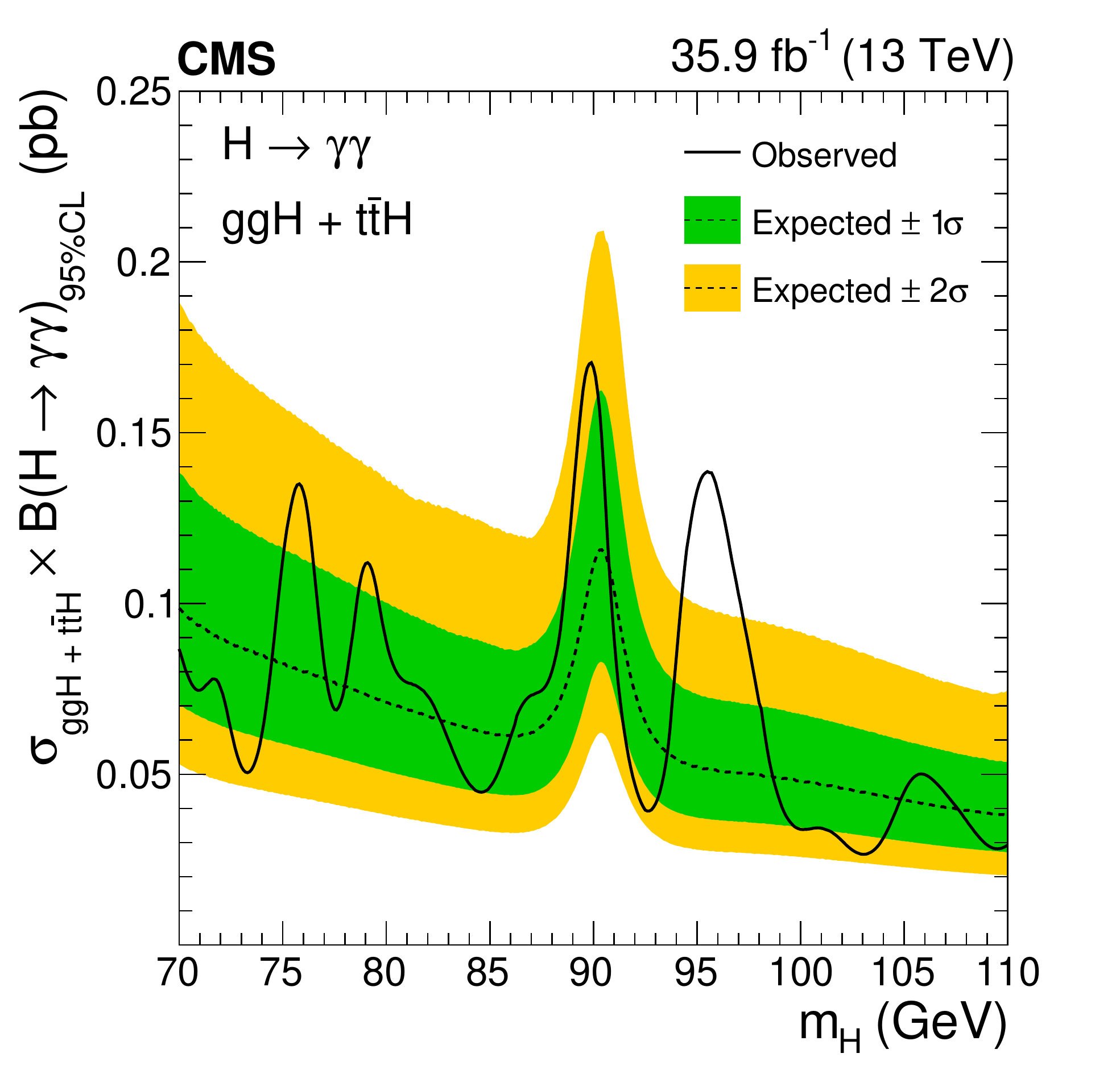}
\includegraphics[width=0.49\textwidth]{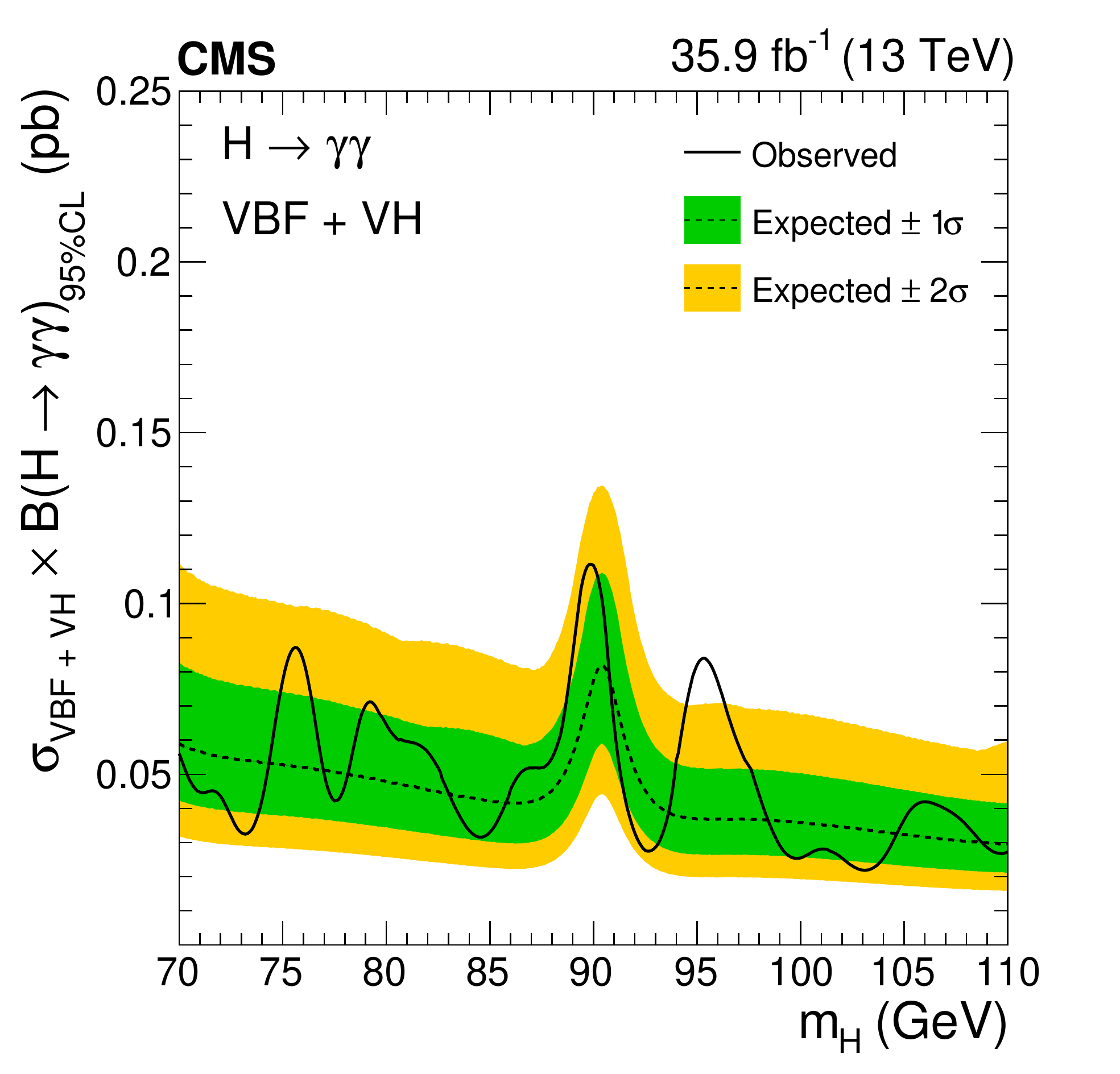}
   \caption{Expected and observed exclusion limits (95\% \CL, in the asymptotic approximation) on the product of the
production cross section
and branching
fraction into two photons for an
additional SM-like Higgs boson,
for the
$\Pg\Pg\PH$ plus
$\ttbar\PH$ (left) and VBF plus $\mathrm{V}\PH$
(right) processes, from the analysis of the 8 (top) and 13 (bottom)\TeV
data. The inner
and
outer
bands indicate the regions containing the distribution of limits located within $\pm$1 and 2$\sigma$,
respectively, of the
expectation under the background-only hypothesis.}
     \label{fig:8TeVLimitSplitAbs}
   \end{figure*}

 In addition, the expected and observed 95\%
\CL upper limits for the
$\Pg\Pg\PH$ plus $\ttbar\PH$
processes and for the VBF plus $\mathrm{V}\PH$ processes are shown in
Fig.~\ref{fig:8TeVLimitSplitAbs}
for each of the 8 and 13\TeV data sets.
The production processes, in each case,
are
combined assuming
relative proportions
as predicted by the SM.

\begin{figure}[htb]\centering
    \includegraphics[width=\cmsFigWidth]{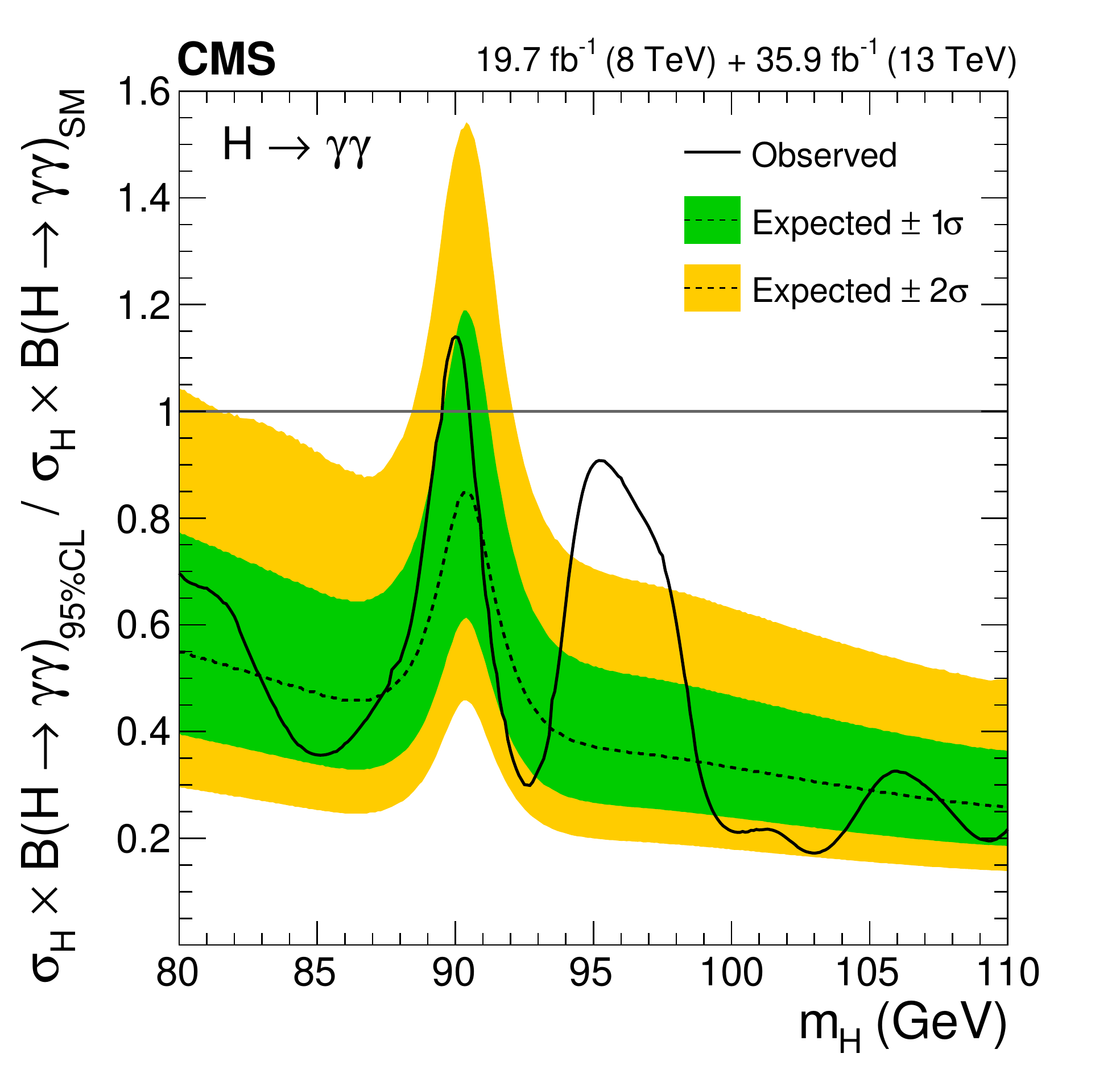}
    \caption{
      \label{fig:8and13TeVNormLimit}
       Expected and observed exclusion limits (95\% \CL, in the asymptotic approximation) on the product of the production cross section
and branching
fraction into two photons for an
additional Higgs boson, relative to the
expected
SM-like
value,
from the analysis of the 8 and 13\TeV data. The inner
and
outer
bands indicate the regions containing the distribution of limits located within $\pm$1 and 2$\sigma$,
respectively, of the
expectation under the background-only hypothesis.
    }
\end{figure}

\begin{figure}[htb]\centering
   \includegraphics[width=\cmsFigWidth]{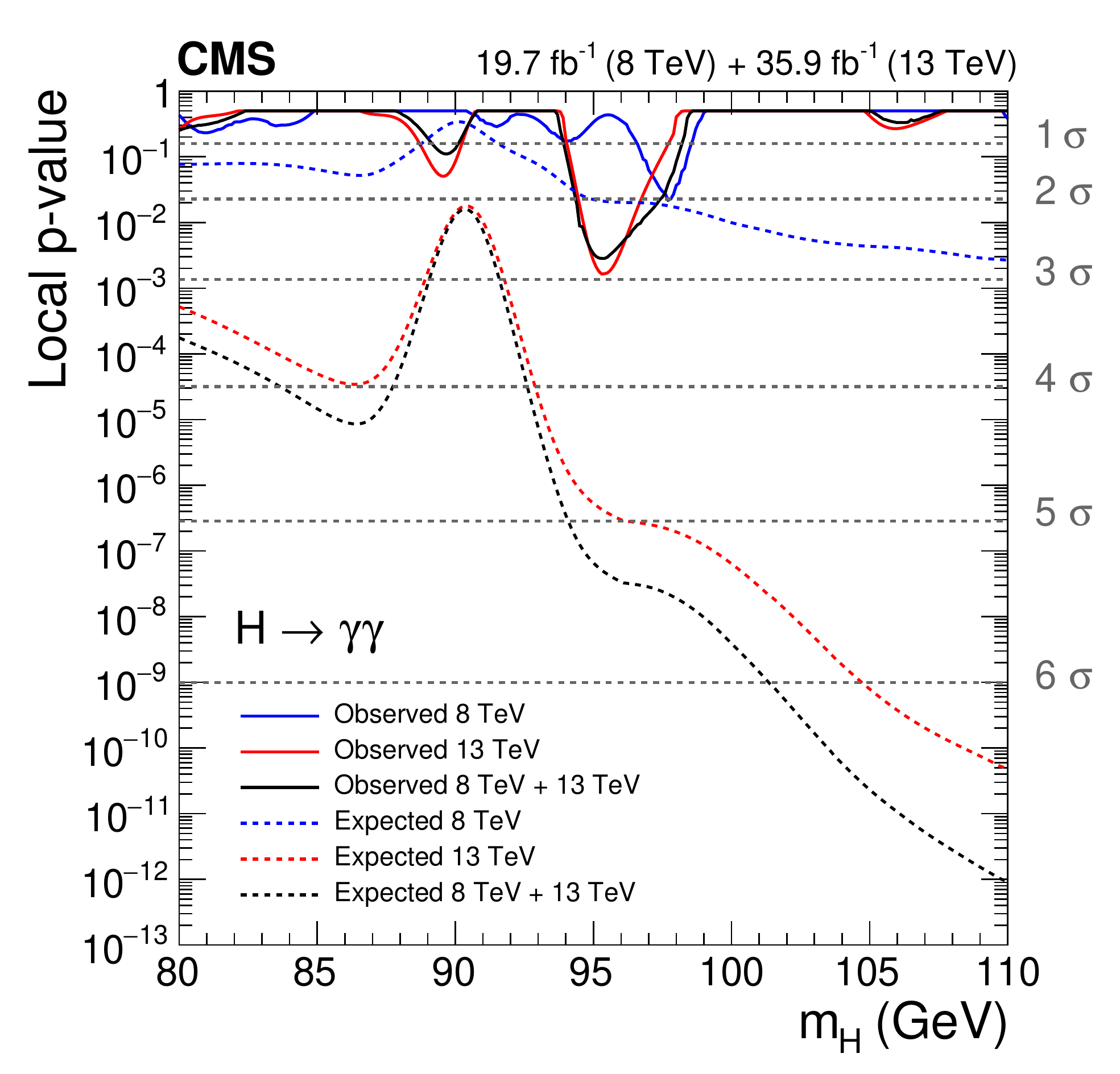}
    \caption{
      \label{fig:decomposedPvalue}
Expected and observed local $p$-values as a function of $\mH$ for the 8 and 13\TeV data and their combination (solid curves) plotted
together with the relevant expectations for an
additional SM-like
Higgs boson
(dotted curves).
    }
\end{figure}

The results from the 8 and 13\TeV data are combined statistically
applying the same methods used to obtain the results from each individual data set, in the diphoton invariant mass range common to the two data sets, $80<\mgg<110\GeV$. All of the experimental systematic uncertainties as well as the theoretical uncertainties
in the signal acceptance due to PDF
variations
are assumed to be uncorrelated between the two data sets. The theoretical uncertainties
in the signal acceptance due to scale variations as well as
in the production cross sections at the center-of-mass energies of 8 and 13\TeV for an
additional SM-like
Higgs boson are assumed to be
fully correlated.
Figure~\ref{fig:8and13TeVNormLimit} shows the expected and observed 95\%
\CL upper limits on the product of the
cross section
and branching
fraction into two photons for an
additional Higgs boson, relative to the SM-like
value from the latest theoretical predictions from the LHC Higgs cross section working group~\cite{deFlorian:2016spz}. No significant excess
with respect to the expected number of background events
is observed. The minimum (maximum) observed upper limit on the product of the production cross section
and branching
fraction normalized to the SM-like
value is 0.17\,(1.13)  corresponding to a
mass hypothesis of 103.0\,(90.0)\GeV.
Figure~\ref{fig:decomposedPvalue} shows the expected and observed local $p$-values as a function of the mass of an
additional SM-like
Higgs boson, calculated with respect to the background-only hypothesis,
from the analyses of the 8 and 13\TeV data, and
from
their combination.  The most significant expected sensitivity occurs at the highest explored
mass hypothesis of
110\GeV with a local expected significance
close to 3$\sigma$ (${>}6\sigma$) for the 8\,(13)\TeV data, while the worst expected significance occurs in
the neighborhood of 90\GeV, where it is
approximately 0.4$\sigma$ (slightly above 2$\sigma$).
For the combination, the most (least) significant expected sensitivity occurs at a mass hypothesis of 110\,(90)\GeV with a local expected significance of approximately 6.8$\sigma$ (slightly above 2.0$\sigma$).
In the case of the 8\TeV data, one excess with approximately 2.0$\sigma$ local significance is observed for a
mass hypothesis of 97.7\GeV. For the 13\TeV data, one excess with approximately 2.90$\sigma$ local (1.47$\sigma$ global) significance is observed for a
mass hypothesis of 95.3\GeV, where the global significance has been calculated
using the method of~\cite{Gross:2010qma}.
In the combination,
an excess with approximately 2.8$\sigma$ local (1.3$\sigma$ global) significance is observed for
a
mass hypothesis of
95.3\GeV.

\section{Summary}

A search for an additional, SM-like,
low-mass Higgs boson
decaying into two photons has been presented. It is based upon data samples corresponding to
integrated luminosities of
19.7
and 35.9\fbinv
collected
at
center-of-mass energies of 8\TeV in 2012 and 13\TeV in 2016, respectively. The search is performed in a mass range between 70 and 110\GeV.
The expected
and observed
95\%
\CL upper limits on the product of the production cross section
and branching
fraction into two photons for an
additional SM-like Higgs boson
as well as the expected
and observed
local $p$-values
are presented.
No significant (${>}3\sigma$) excess
with respect to the expected number of background events
is observed.
The observed upper limit on the product of the production cross section and branching fraction for the 2012\,(2016)
data set ranges from 129\,(161)\unit{fb} to 31\,(26)\unit{fb}.
The statistical combination of the results
from the analyses of the two data sets in the common mass range between 80
and 110\GeV yields an upper limit on the product of the cross section and branching
fraction, normalized to that for a standard model-like Higgs boson, ranging from
0.7 to 0.2, with two notable exceptions: one in the region around the $\PZ$ boson peak,
where the limit rises to 1.1, which may be due to the presence of Drell--Yan dielectron production
where electrons could be misidentified as isolated photons, and a second due to an
observed excess with respect to the standard model prediction, which is maximal for
a mass hypothesis of 95.3\GeV with a local (global) significance of 2.8\,(1.3) standard
deviations.
More data are required to ascertain the origin of this excess.
This is the first
search for new resonances in the diphoton final state in this mass range
based on LHC data at
a
center-of-mass energy of 13\TeV.

\begin{acknowledgments}
We congratulate our colleagues in the CERN accelerator departments for the excellent performance of the LHC and thank the technical and administrative staffs at CERN and at other CMS institutes for their contributions to the success of the CMS effort. In addition, we gratefully acknowledge the computing centers and personnel of the Worldwide LHC Computing Grid for delivering so effectively the computing infrastructure essential to our analyses. Finally, we acknowledge the enduring support for the construction and operation of the LHC and the CMS detector provided by the following funding agencies: BMBWF and FWF (Austria); FNRS and FWO (Belgium); CNPq, CAPES, FAPERJ, FAPERGS, and FAPESP (Brazil); MES (Bulgaria); CERN; CAS, MoST, and NSFC (China); COLCIENCIAS (Colombia); MSES and CSF (Croatia); RPF (Cyprus); SENESCYT (Ecuador); MoER, ERC IUT, and ERDF (Estonia); Academy of Finland, MEC, and HIP (Finland); CEA and CNRS/IN2P3 (France); BMBF, DFG, and HGF (Germany); GSRT (Greece); NKFIA (Hungary); DAE and DST (India); IPM (Iran); SFI (Ireland); INFN (Italy); MSIP and NRF (Republic of Korea); MES (Latvia); LAS (Lithuania); MOE and UM (Malaysia); BUAP, CINVESTAV, CONACYT, LNS, SEP, and UASLP-FAI (Mexico); MOS (Montenegro); MBIE (New Zealand); PAEC (Pakistan); MSHE and NSC (Poland); FCT (Portugal); JINR (Dubna); MON, RosAtom, RAS, RFBR, and NRC KI (Russia); MESTD (Serbia); SEIDI, CPAN, PCTI, and FEDER (Spain); MOSTR (Sri Lanka); Swiss Funding Agencies (Switzerland); MST (Taipei); ThEPCenter, IPST, STAR, and NSTDA (Thailand); TUBITAK and TAEK (Turkey); NASU and SFFR (Ukraine); STFC (United Kingdom); DOE and NSF (USA).

\hyphenation{Rachada-pisek} Individuals have received support from the Indo-French Network in High Energy Physics financed by the Indo-French Center for the Promotion of Advanced Research (CEFIPRA/IFCPAR), the Marie-Curie program and the European Research Council and Horizon 2020 Grant, contract No. 675440 (European Union); the Leventis Foundation; the A. P. Sloan Foundation; the Alexander von Humboldt Foundation; the Belgian Federal Science Policy Office; the Fonds pour la Formation \`a la Recherche dans l'Industrie et dans l'Agriculture (FRIA-Belgium); the Agentschap voor Innovatie door Wetenschap en Technologie (IWT-Belgium); the F.R.S.-FNRS and FWO (Belgium) under the ``Excellence of Science - EOS" - be.h project n. 30820817; the Ministry of Education, Youth and Sports (MEYS) of the Czech Republic; the Lend\"ulet (``Momentum") Program and the J\'anos Bolyai Research Scholarship of the Hungarian Academy of Sciences, the New National Excellence Program \'UNKP, the NKFIA research grants 123842, 123959, 124845, 124850 and 125105 (Hungary); the Council of Science and Industrial Research, India; the HOMING PLUS program of the Foundation for Polish Science, cofinanced from European Union, Regional Development Fund, the Mobility Plus program of the Ministry of Science and Higher Education, the National Science Center (Poland), contracts Harmonia 2014/14/M/ST2/00428, Opus 2014/13/B/ST2/02543, 2014/15/B/ST2/03998, and 2015/19/B/ST2/02861, Sonata-bis 2012/07/E/ST2/01406; the National Priorities Research Program by Qatar National Research Fund; the Programa Estatal de Fomento de la Investigaci{\'o}n Cient{\'i}fica y T{\'e}cnica de Excelencia Mar\'{\i}a de Maeztu, grant MDM-2015-0509 and the Programa Severo Ochoa del Principado de Asturias; the Thalis and Aristeia programs cofinanced by EU-ESF and the Greek NSRF; the Rachadapisek Sompot Fund for Postdoctoral Fellowship, Chulalongkorn University and the Chulalongkorn Academic into Its 2nd Century Project Advancement Project (Thailand); the Welch Foundation, contract C-1845; and the Weston Havens Foundation (USA).
\end{acknowledgments}
\bibliography{auto_generated}

\cleardoublepage \appendix\section{The CMS Collaboration \label{app:collab}}\begin{sloppypar}\hyphenpenalty=5000\widowpenalty=500\clubpenalty=5000\vskip\cmsinstskip
\textbf{Yerevan Physics Institute, Yerevan, Armenia}\\*[0pt]
A.M.~Sirunyan, A.~Tumasyan
\vskip\cmsinstskip
\textbf{Institut f\"{u}r Hochenergiephysik, Wien, Austria}\\*[0pt]
W.~Adam, F.~Ambrogi, E.~Asilar, T.~Bergauer, J.~Brandstetter, E.~Brondolin, M.~Dragicevic, J.~Er\"{o}, A.~Escalante~Del~Valle, M.~Flechl, M.~Friedl, R.~Fr\"{u}hwirth\cmsAuthorMark{1}, V.M.~Ghete, J.~Hrubec, M.~Jeitler\cmsAuthorMark{1}, N.~Krammer, I.~Kr\"{a}tschmer, D.~Liko, T.~Madlener, I.~Mikulec, N.~Rad, H.~Rohringer, J.~Schieck\cmsAuthorMark{1}, R.~Sch\"{o}fbeck, M.~Spanring, D.~Spitzbart, A.~Taurok, W.~Waltenberger, J.~Wittmann, C.-E.~Wulz\cmsAuthorMark{1}, M.~Zarucki
% ...
\vskip\cmsinstskip
% ...
1:  Also at Vienna University of Technology, Vienna, Austria\\
% ...
\end{sloppypar}
\end{document}